\documentclass[12pt]{article}
\usepackage[T1]{fontenc}
\usepackage{lmodern}   
\usepackage{ragged2e}
\usepackage{microtype}
\usepackage{geometry}
\usepackage{bm}
\usepackage{babel}
\usepackage{float}
\usepackage{rotfloat} 
\usepackage{tablefootnote}
\usepackage{mathtools}

\usepackage{algorithm}
\usepackage[noend]{algpseudocode}
\usepackage{algorithmicx}
\usepackage{setspace}
\usepackage{amsfonts}
\usepackage{natbib}
\usepackage{subcaption}
\usepackage[section]{placeins}
\usepackage{longtable}
\usepackage{bbold}
\usepackage{hyperref}
\usepackage{comment}
\usepackage[table]{xcolor}

\hypersetup{
  colorlinks=true,
  linkcolor=purple,
  filecolor=magenta,   
  urlcolor=purple,
  citecolor=blue,
}
\usepackage{titlesec}
\usepackage{xr}
\usepackage{cleveref}

\usepackage{amsmath}
\usepackage{graphicx,psfrag,epsf,epstopdf}
\usepackage{enumerate}
\usepackage{url} 

\usepackage{amssymb}
\usepackage{amsthm}
\usepackage{color}
\usepackage{caption}
\usepackage{booktabs}
\usepackage{bbm}
\usepackage{multicol}
\usepackage{multirow}
\usepackage{rotating}
\usepackage{pdflscape}

\def\bmath0{{\boldmath 0}}
\def\bm1{{\boldsymbol 1}}

\def\bmu{{\boldsymbol u}}

\def\bmx{{\boldsymbol x}}
\def\bmy{{\boldsymbol y}}

\def\bmp{{\boldsymbol p}}

\def\bmz{{\boldsymbol z}}

\def\bmX{{\boldsymbol X}}

\def\bmP{{\boldsymbol P}}
\def\bmU{{\boldsymbol U}}

\newtheorem{definition}{Definition}
\newtheorem{corollary}{Corollary}

   \def\0{\boldsymbol 0}

\begin{document}

\title{\LARGE\bf Massive parallelization of projection-based depths}

\author{Leonardo Leone \\ {\small CREST, CNRS, LTCI, T{\'e}l{\'e}com Paris, Institut Polytechnique de Paris}\\
Pavlo Mozharovskyi \\ {\small LTCI, T{\'e}l{\'e}com Paris, Institut Polytechnique de Paris}\\
David Bounie \\ {\small CREST, CNRS, T{\'e}l{\'e}com Paris, Institut Polytechnique de Paris}}
    
\date{June 09, 2025}
\maketitle

\vspace{5ex}

\begin{abstract}
This article introduces a novel methodology for the massive parallelization of projection-based depths, addressing the computational challenges of data depth in high-dimensional spaces. 
We propose an algorithmic framework based on Refined Random Search (RRS) and demonstrate significant speedup (up to 7,000 times faster) on GPUs. Empirical results on synthetic data show improved precision and reduced runtime, making the method suitable for large-scale applications. The RRS algorithm (and other depth functions) are available in the \texttt{\textbf{Python}}\textbf{\textit{-library}} {\bf \href{https://data-depth.github.io/}{data-depth}}  with ready-to-use tools to implement and to build upon this work.
\end{abstract}



\newpage
\section{Introduction}

The concept of data depth has become a cornerstone in multivariate statistics, offering a non-parametric measure of centrality and outlyingness in high-dimensional spaces. Introduced by \cite{tukey1975mathematics}, data depth functions quantify the centrality of a point within a data set, providing a robust alternative to traditional distance-based metrics. However, the computational complexity of calculating data depth, especially in high-dimensional settings, has hindered its practical application. Current work addresses these challenges by introducing a novel methodology for the massive parallelization of projection-based depths, significantly enhancing computational efficiency and scalability.

\medskip

Projection-based depths, such as halfspace depth, projection depth, and asymmetric projection depth, are particularly well-suited for high-dimensional data analysis due to their affine invariance and robustness. These depths satisfy the weak projection property, allowing multivariate depths to be computed as the minimum of univariate depths across projections in all directions. However, exploring all possible directions in high-dimensional spaces is computationally unfeasible, necessitating optimization routines like the Refined Random Search (RRS). While RRS has shown promise, it is often limited by sequential computation, failing to fully leverage modern parallel computing architectures.

\medskip

In this article, we propose an algorithmic framework that harnesses GPU acceleration to achieve massive parallelization of projection-based depths. Our approach builds upon the RRS algorithm and extends it by decomposing the computation into highly parallelizable tasks suited for GPU architecture. First, we formalize characteristic time equations for both our parallel approach ($T_G$) and the standard approach ($T_C$), and derive speedup equations ($S_p$) for RRS inspired by Amdahl's law~\citep{amdahl2007validity}. The theoretical speedup $S_p$ is expressed as a quotient involving $\lambda \cdot d$, where $\lambda$ is the ratio of CPU to GPU frequency ($\lambda = \frac{f_{CPU}}{f_{GPU}}$), $d$ is the dimension, and $g$ is the number of available GPU cores. This reevaluation of Amdahl's law highlights speedup limitations due to core speed, core quantity, and sequential operations.

\medskip

Second, we decompose the computation of projection-based depths into parallelizable tasks, achieving significant speedups of up to 7,000 times faster than state-of-the-art methods, particularly when implemented on GPUs like the NVIDIA GeForce RTX 4080 (see \citep{dyckerhoff2016exact,fojtik2023exact}, and for projection depth \citep{zuo2011exact}). The key innovation lies in the parallelization of the three main operations involved in depth computation: direction generation, data projection, and univariate depth calculation. By breaking these operations into simple, \textit{embarrassingly parallel} tasks, we achieve efficient use of GPU resources, resulting in dramatic reductions in runtime without sacrificing precision.

\medskip

Third, we provide insights into hyperparameter fine-tuning, such as the selection of spherical cap shrinkage, the number of refinements, and directions, to optimize convergence and runtime. These findings ensure the robustness and adaptability of our method across different data sets and dimensionalities. The center-outward ordering precision, evaluated through ranking correlation techniques, confirms the reliability of our approach in high-dimensional data analysis.

\medskip

Our contributions are multifold. First, we introduce a novel algorithmic framework for the massive parallelization of projection-based depths, achieving significant speedups on GPUs. Second, we provide a comprehensive analysis of the computational challenges associated with data depth and demonstrate the effectiveness of our approach in overcoming these challenges. Third, we explore hyperparameter fine-tuning for the RRS algorithm, providing insights into optimal parameter selection for large-scale scenarios. Finally, we conduct a thorough runtime analysis, demonstrating the empirical speedup achieved by our method and validating our theoretical model. These contributions not only advance the state-of-the-art in data depth computation but also pave the way for broader adoption of depth-based methods in practical applications. By overcoming the computational bottlenecks, we enable researchers and practitioners to leverage the full potential of data depth in large-scale, high-dimensional settings, opening new avenues for robust statistical analysis.

\medskip

The remainder of the paper is organized as follows. Section~\ref{sec:DD_tech} provides a detailed background on data depth, focusing on projection-based depths and their computational challenges. Section~\ref{sec:extPara} introduces our methodology for massive parallelization, detailing the parallelization of each operation and the theoretical underpinnings of our approach. Section~\ref{sec:AlgoExpl} presents the complete massively-parallel algorithm, including the integration of parallel directions generation, projection, and depth computation. In Section~\ref{sec:fineTun}, we explore hyperparameter fine-tuning for the RRS algorithm, providing insights into optimal parameter selection for large-scale scenarios. Section~\ref{sec:timeStudy} conducts a thorough runtime analysis, demonstrating the empirical speedup achieved by our method and validating our theoretical models. Finally, Section~\ref{sec:rankOrder} evaluates the precision of our approach through center-outward ordering and depth benchmarking, confirming its effectiveness in high-dimensional data analysis. Section~\ref{sec:Conclusion} concludes.

\section{Background on data depth} \label{sec:DD_tech}

In this section, we review the most relevant aspects of data depth from an optimization perspective. For simplicity, we focus on the empirical version of data depth throughout this article. Following Tukey's seminal idea \citep{tukey1975mathematics}, data depth is a statistical function that measures the degree to which an arbitrary observation belongs to a data set. Consider a data set $\bmX=\{\bmx_1,...,\bmx_n\}\subset\mathbb{R}^d$ (using set notation for convenience, though ties are generally not excluded) and a point $\bmz\in\mathbb{R}^d$.
\begin{definition}\label{def:depth}
    \emph{Data depth} is a function
    \begin{equation*}
        D\,:\,\mathbb{R}^d \times \mathbb{R}^{n \times d} \rightarrow [0, 1]\,,\,(\bmz,\bmX)\mapsto D(\bmz|\bmX)\,,
    \end{equation*}
    which satisfies the properties of affine invariance, vanishing at infinity, monotonicity on rays originating from the deepest point, and upper-semicontinuity.
\end{definition}

For a comprehensive and formal discussion on data depth and its axiomatics, we refer the reader to \cite{zuo2000general} and \cite{mosler2022choosing}. In this article,
we focus on a property that is particularly relevant to our current analysis: \emph{the weak projection property}. Let $\mathbb{S}^{d-1}=\{\bmx\in\mathbb{R}^d\,:\,\|\bmx\|=1\}$ denote the unit sphere and $^\top$ be the transposition operator.

\begin{definition}[\cite{dyckerhoff2004data}]\label{def:projprop}
 The function $D$ from Definition~\ref{def:depth} satisfies the \emph{weak projection property} if for any $\bmz\in\mathbb{R}^d,\bmX\subset\mathbb{R}^d$, it holds:
    \begin{equation*}
        D(\bmz|\bmX) = \inf_{\bmu\in\mathbb{S}^{d-1}} D(\bmz^\top\bmu|\bmX^\top\bmu)\,.
    \end{equation*}
\end{definition}

In the following, we will omit  the term ``weak'', and simply refer to depths that satisfy the \emph{projection property}. These include Mahalanobis~\citep{mahalanobis2018generalized}, halfspace~\citep{tukey1975mathematics}, zonoid~\citep{koshevoy1997zonoid}, projection~\citep{zuo2000general}, asymmetric projection~\citep{dyckerhoff2004data}, (continuous) expected convex hull~\citep{Cascos07}, geometrical~\citep{DyckerhoffM11} depths. Definition~\ref{def:projprop} leads to the following computationally important corollary:
\begin{corollary}
    Let $\bmu_1,...,\bmu_k\subset\mathbb{S}^{d-1}$, then:
    \begin{equation*}
        D(\bmz|\bmX) \le \min_{j=1,...,k} D(\bmz^\top\bmu_k|\bmX^\top\bmu_k)\,.
    \end{equation*}
\end{corollary}

In other words, multivariate data depths can be computed as the minimum of univariate depths across projections in all directions. Since exploring all possible directions (an infinite set) is computationally infeasible, a finite subset must be examined using an optimization routine. Within this routine, only univariate depths need to be computed.

\medskip

For any $\bmu\in\mathbb{S}^{d-1}$, let $\mathbb{R}\ni y=\bmz^\top\bmu$ and $\mathbb{R}^n\ni\bmy=\bmX^\top\bmu\equiv\{\bmx_1^\top\bmu,...,\bmx_n^\top\bmu\}$, such that a depth satisfying the projection property can be written as $\inf D(y|\bmy)$. A depth satisfying the projection property can then be expressed as $\inf D(y|\mathbf{y})$. 

\medskip

Among the various depths that satisfy the projection property, we will focus on three robust and geometrically affine-invariant notions throughout this article:
\begin{itemize}
    \item \emph{Halfspace depth}:
    \begin{equation*}
        D_H(y|\bmy) = \frac{1}{n}\min \bigl\{ \#(\bmy \le y) , \#(\bmy \ge y) \bigr\}\,,
    \end{equation*}
    where $\#$ denotes the cardinal of a set, with the vector-scalar operation applicable element-wise.
    \item \emph{Projection depth}:
    \begin{equation*}
        D_P(y|\bmy) = \Bigl(1 + \frac{|y - \text{med}(\bmy)|}{\text{MAD}(\bmy)}\Bigr)^{-1}\,,
    \end{equation*}
    where $\text{med}(\bmy)$ denotes the (usual) univariate median and $\text{MAD}(\bmy) = \text{med}\bigl(|\bmy-\text{med}(\bmy)|\bigr)$ stands for the median absolute deviation for the median.
    \item \emph{Asymmetric projection depth}:
    \begin{equation*}
        D_{AP}(y|\bmy) = \Bigl(1 + \frac{\bigl(y - \text{med}(\bmy)\bigr)_+}{\text{MAD}_+(\bmy)}\Bigr)^{-1}\,,
    \end{equation*}
    where $(a)_+=\max \{a, 0\}$ and $\text{MAD}_+(\bmy)$ is the median of the positive deviations from the median.
\end{itemize}

It is noteworthy that all three of the aforementioned univariate notions are computable with the best achievable complexity of $\mathcal{O}(n)$. However, implementation details may reveal important differences, as we will discuss in Section \ref{sec:AlgoExpl}.

\section{Massive Parallel Computation of Data Depth}\label{sec:extPara}

We propose a viable methodology to compute the depth of a point considering a large number of directions and samples in the data set. 

\medskip

A projection-based notion follows three (main) operations to compute the final depth:

\begin{enumerate}
    \item \emph{Generating} $k$ directions in $\mathbb{S}^{d-1}$.
    \item \emph{Projecting} the data onto these directions.
    \item \emph{Computing} the univariate depth within each projection.
\end{enumerate}

\begin{figure}
    \centering
    \includegraphics[width=150mm]{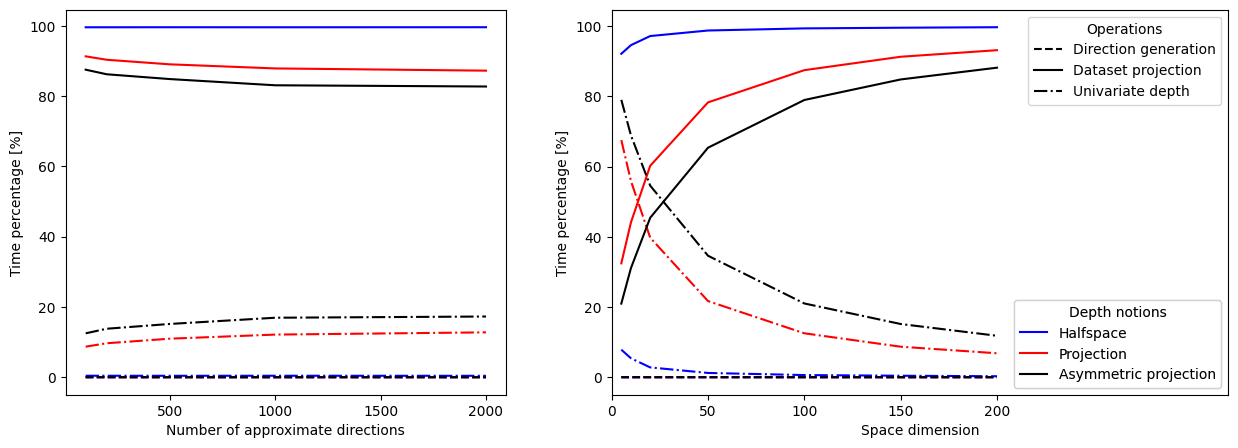} 
    \caption{The time breakdown for computing projection depth on CPU without parallelization is illustrated as follows: \textcolor{blue}{blue lines} represent the time spent generating directions in the unit sphere, following a normal distribution; \textcolor{red}{red lines} represent the time spent projecting onto these directions; \textbf{black lines} represent the time spent computing the univariate (projection) depth. Left graph: Displays the percentage of time spent on each operation, with $n$ (number of points) fixed at 1,000 and $d$ (space dimension) to 150. Right graph: Displays the percentage of time spent on each operation with respect to the space dimension $d$, with $n$ and $k$ (number of directions) fixed at 1,000.}
    \label{fig:BdTimeCpp}
\end{figure}

The time importance of each operation is illustrated in Figure \ref{fig:BdTimeCpp}. Here, directions are randomly generated following a uniform distribution in $\mathbb{S}^{d-1}$ (\textit{i.e.}, without refinements). The univariate depth is computed using three depth notions: $D_H$ in blue lines, $D_P$ in red lines, and $D_{AP}$ in black lines.

\medskip

The left graph shows the time percentage of each operation with respect to the number of directions ($k$), with the space dimension $d$ fixed at 150 and the data set $\bmX$ containing 1,000 samples. We note that the time percentage remains nearly constant for all $k$, as the time for both main operations (direction generation and projection) grows proportionally with the number of directions. Overall, projection operations dominate the runtime, being the most time-consuming step. 

\medskip

The right graph, using 1,000 directions and a data set with 1,000 samples, examines the effect of the space dimension on time percentage. Here, $D_P$ and $D_{AP}$ show a greater emphasis on univariate depth computation, as these notions rely on (practically) more time-consuming operations like the median. In contrast, $D_H$ relies solely on value comparisons, making it less computationally intensive in this context.

\medskip

The time complexity of these three operations depends on the number of points $n$, directions $k$, and space dimensions $d$, reflecting their computational significance.

\begin{enumerate}
    \item Generating directions: This operation has a time complexity of $\mathcal{O}(kd)$ and is the least computationally intensive among the three, as it involves generating $k$ random directions in a $d$-dimensional space.

    \item Projecting data: This step has a time complexity of $\mathcal{O}(knd)$, which grows with the space dimension $d$ due to the tensor multiplications required to project the data onto the generated directions. This operation is more computationally demanding than generating directions.

    \item Computing univariate depths: The time complexity of this operation depends on the chosen depth notion. For instance, in the case of projection depth ($D_P$), the complexity is $\mathcal{O}(kn)$. For a generic depth notion, the complexity is denoted as $\mathcal{O}(\mathcal{D})$. This step can be computationally significant, especially for depth notions that involve more complex calculations, such as the median.
\end{enumerate}

\medskip

With a resulting time complexity of $\mathcal{O}(kd)\cup\mathcal{O}(knd)\cup\mathcal{O}(\mathcal{D})$, the objective of this article is to develop a method for massive parallelization of the rate-determining operation, $\mathcal{O}(knd)$, using an optimization routine called Refined Random Search (RRS). This approach aims to significantly reduce computational time while maintaining the precision of the algorithm. 

\medskip

Several optimization methods have been used to identify the direction that minimizes the depth of a point, including (best performing) RRS, coordinate descent, and the Nelder-Mead method \citep{dyckerhoff2021approximate}. Among these methods, we selected RRS due to its balance of precision and parallelization potential, making it the most suitable for constructing our framework methodology. This algorithm is further explored in Section \ref{sec:AlgoExpl}. 

\medskip

Our approach leverages the complete parallel computation of depth, including both projection and univariate depth calculation, for a point with respect to different directions within the same refinement. By focusing on data projection, this approach allows our framework to decompose $\mathcal{O}(knd)$ into synchronous, parallel calculations. 

\medskip

Let us introduce our approach with the following model. All mentioned operations result in a characteristic time $T_C$, where each operation has an associated cost and respective time complexity. We denote the total number of directions per refinement as $m$, where $m=\left\lceil{\frac{k}{r}}\right\rceil$, $T_C$ can be written as: 
\begin{equation}
\label{eqn:CPU_time}
T_{C}=C_{c}+r\cdot \left(C_{rv}\cdot m\cdot d+C_{p}\cdot m\cdot n\cdot d+C_{d_{(1)}}\cdot \mathcal{D}\right), 
\end{equation}
where $C_{c}$ is a constant computational cost, $C_{rv}$ is the cost of generating random directions, $C_{p}$ is associated with projecting data, $C_{d_{(1)}}$ represents the additional multiplicative factor of the univariate depth computation. These operations have distinct computational characteristics and are parallelized in different ways. Our goal is to break down each of these steps into multiple simple tasks suitable for massive parallel (\textit{e.g.}, with GPU) computation.

\medskip

Most of these operations can be reduced to tensor multiplication(s), floating-point operations (FLOPS), and conditional statements. In parallel computing, these operations are referred to as \textit{embarrassingly parallel}, as most of their steps can be performed concurrently \citep{navarro2014survey}. In $g$-massive parallel computation (\textit{e.g.}, in a GPU with $g$ computing units):

\begin{itemize}
\item generating random directions can be performed in $\left\lceil{\frac{m\cdot d}{g}}\right\rceil$ steps,
\item depth computation can be completed in $\left\lceil\frac{\mathcal{D}}{g}\right\rceil$ steps, 
\item projecting data can be executed in $\left\lceil{\frac{m\cdot n\cdot d}{g}}\right\rceil$ steps.
\end{itemize}

Let us denote with the term $\left\lceil{\frac{d}{d(m,n)_{max}}}\right\rceil$ the breakdown of the inner product ($\bmx\bmu^\top$). Depending on the size of the vectors, vector multiplication might be further divided into smaller kernels to accommodate memory usage: 
$$ \boldsymbol{x} \boldsymbol{u}^\top=\sum_{i=1}^{d} x_i\cdot u_i = \sum_{i=1}^{p} x_i\cdot u_i+\sum_{j=p+1}^{d} x_j\cdot u_j.$$ 

Further, let $\lambda$ denote the ratio between the massive parallel frequency (\textit{e.g.}, a GPU core frequency) and the basic (non-parallel) frequency (\textit{e.g.}, a CPU core frequency) to be considered in the parallel time equation. With the frequency ratio $\lambda=\frac{f_{CPU}}{f_{GPU}}$, characteristic time for our parallel approach $T_{G}$ can be written as: 
\begin{equation}
\label{eqn:GPU_time}
T_{G}=C_{c}+r\cdot\lambda\cdot \Biggl(C_{rv}\cdot \left\lceil{\frac{m\cdot d}{g}}\right\rceil+C_{p}\cdot \left\lceil{\frac{d}{d_{max(m,n)}}}\right\rceil\cdot \left\lceil{\frac{m\cdot n}{g}}\right\rceil+C_{d_{(1)}}\cdot \left\lceil{\frac{\mathcal{D}}{g}}\right\rceil\Biggr).
\end{equation}

\medskip

Following Amdahl's law for parallel speedup, we can construct a theoretical time ratio to model the behavior of our approach in a system with $g$ massively-parallel (\textit{e.g.}, GPU) cores \citep{amdahl2007validity, gustafson1988reevaluating, marowka2012extending}. Speedup ($S_p$) is formalized as the ratio between characteristic times, expressed as: 
\begin{equation}
S_p=\frac{T_C}{T_G}.
\label{eqn:Sp}
\end{equation}

After formalizing these equations, we now propose an optimized algorithmic framework for massively-parallel depth parallelization. 

\section{Algorithmic framework for massive parallelization}
\label{sec:AlgoExpl}

Refined Random Search algorithm combines purely random approximation, using independent directions uniformly distributed over the unit sphere, with an optimization element. Specifically, this optimization routine consists of three steps, repeated within each refinement. A brief explanation of the main operation is provided in Algorithm~\ref{alg:RRS}. For more details on the RRS algorithm, see \cite{dyckerhoff2021approximate} and Section \ref{Sec:RandSearch} in the Supplementary Material. 

\medskip

\begin{algorithm}[!h]
\footnotesize
\caption{Classical refined random search}

\textbf{Input:} $\bmX\in\mathbb{R}^{n\times d}$, $\bmz\in\mathbb{R}^{d}$, $k$, $r$, $\alpha$.

\textbf{Step 1.} Set $\bmp=\boldsymbol{\mathrm{e}}_1$, $D_{min}=1$, $m=\lceil\frac{k}{r}\rceil$, $\epsilon=\frac{\pi}{2}$.

\textbf{Step 2.} Loop for each refinement: For $l=1$ to $r$:
\begin{enumerate}
    \item[] \textbf{Step 3.} Minimize depth in the $\epsilon$-region: For $j=1$ to $m$:
    \begin{itemize}
        \item[(a)] \textbf{Generating} direction: $\mathbb{R}^d\ni\bmu_j\sim\mathcal{U}(\mathbb{S}^{d-1})$ inside an $\epsilon$-region around $\bmp$.
        \item[(b)] \textbf{Projecting} the data: $\bmy_j=\bmX^\top\bmu_j$ and $y_j=\bmz^\top\bmu_j$.
        \item[(c)] \textbf{Computing} univariate depth: $D_j=D(y_j|\bmy_j)$.
    \end{itemize}
    \item[] \textbf{Step 4.} Minimal depth $D_{l}=\min_{j\in\{1,...,m\}}D_j$ and its respective direction $\boldsymbol{u_{l}}$.
    \item[] \textbf{Step 5.} Reevaluate depth: if $D_l<D_{min}$ update values:
    \begin{itemize}
        \item[(a)] Update minimal depth $D_{min}=D_l$.
        \item[(b)] Update pole $\boldsymbol{p}=\boldsymbol{u}_l$.
    \end{itemize}
    \item[] \textbf{Step 6.} Reduce the size of the region $\epsilon=\epsilon\cdot\alpha$.
\end{enumerate}

\textbf{Output:} $D_{min}$.

\label{alg:RRS}
\end{algorithm}

Step~3 of Algorithm~\ref{alg:RRS} presents the pseudo-code for the three main operations performed within each refinement: generation, projection, and computing of univariate depth. At the end of each refinement, a new pole $\bmp$ is selected---the direction with the minimal computed depth in the current refinement step. This selection reduces the search region by an angle $\alpha$, a process known as spherical cap shrinkage.

\medskip

\begin{figure}[!h]
    \centering
    \includegraphics[width=100mm]{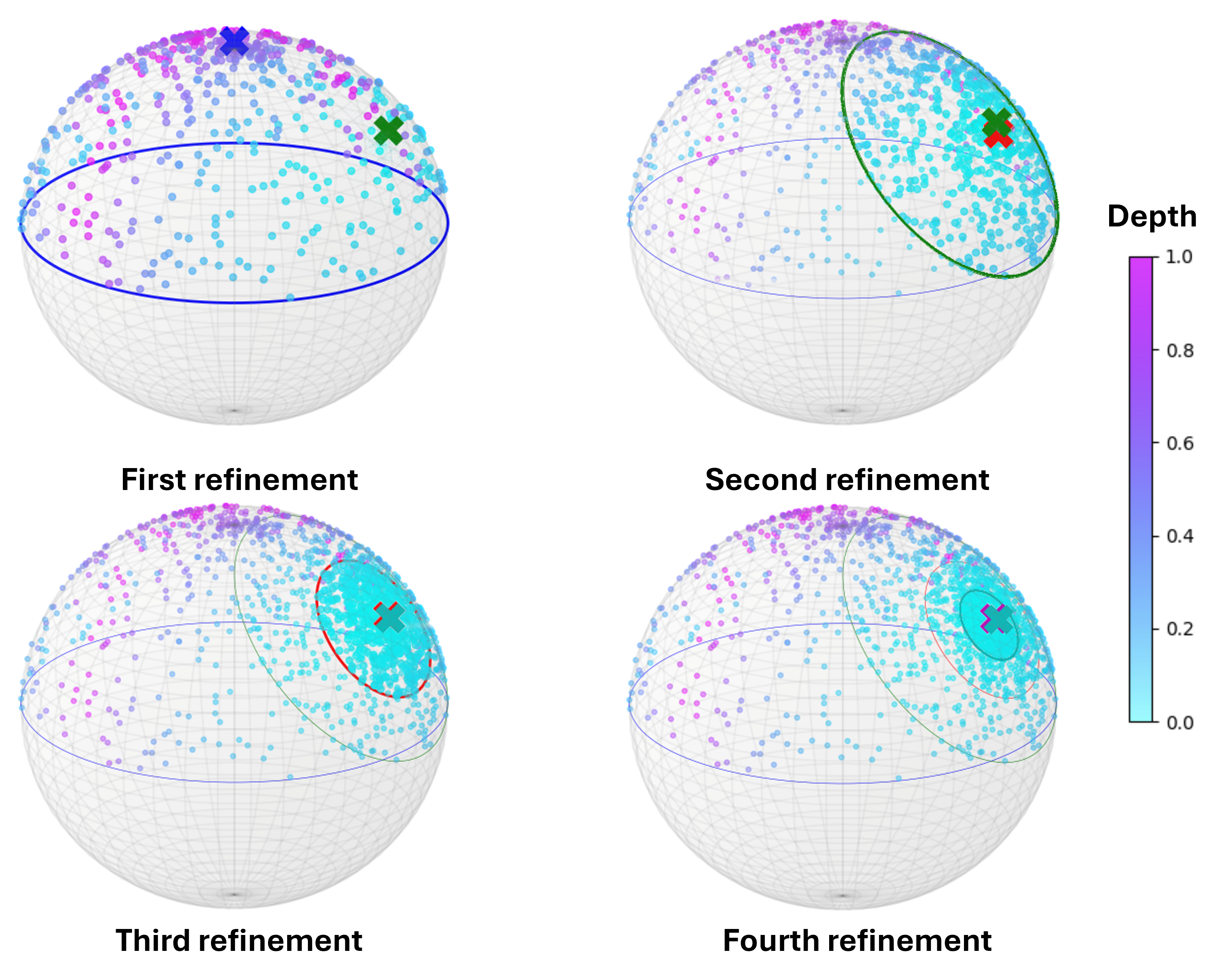}
    \caption{RRS illustration. Points marked with "X" indicate poles, round points represent random directions, and solid lines indicate the limits of the search region. The color of the round points shows the local $D_P$. The pole with the same color as the boundary is the center of the current region while the pole with a different color is the center of the following refinement.}
    \label{fig:RefSeq}
\end{figure}

An illustration of the RRS algorithm for dimension $d = 3$ is shown in Figure~\ref{fig:RefSeq}, where four refinements are necessary to find the minimal depth. In this figure, each point represents a unique direction $\in\mathbb{S}^{d-1}$, with colors' gamma indicating the depth values. Points marked with a cross correspond to the poles, and solid lines represent the borders of the respective spherical caps. Poles and caps can be matched by color, while the cross of the ``new'' color represents the new pole.

A crucial observation enables massive parallelism: while each refinement depends on the previous one, the three operations within the same refinement -- generation, projection, and computation -- are entirely independent for each direction and can thus be run in parallel. Given that the number of random directions within the same refinement step can be substantial, this allows for dramatic acceleration (see the above Section~\ref{sec:extPara} for more details, \textit{cf.} $T_C$ in~\eqref{eqn:CPU_time} and $T_G$ in~\eqref{eqn:GPU_time}).

\subsection{A parallel perspective on the Refined Random Search algorithm}\label{ssec:ParallelRefRand}

The goal is not merely to identify which parts of Algorithm~\ref{alg:RRS} should be parallelized, but to develop a viable algorithm with massive parallelism. To achieve this, we  classify the operations into sequential and parallel components. As illustrated in Figure~\ref{fig:BdTimeCpp}, during purely sequential execution, the most time-consuming operation (with time complexity $\mathcal{O}(knd)$, namely projection) is highly parallelizable within each refinement. In this part, we focus on parallelizing the projection operation. The comprehensive algorithm, including the parallelization of further operations, is discussed in the following Subsection~\ref{ssec:entireAlg}.

\medskip

Let $\bmU=(\bmu_1,...,\bmu_m)^\top$ be a $m \times d$ matrix consisting of directions $\mathbb{R}^d\ni\bmu_1,...,\bmu_m\in\mathbb{S}^{d-1}$ (as rows). With the data set denoted by $\bmX$, the operation to parallelize is the matrix multiplication $\bmP=\bmU\bmX^\top$, where $\bmP$ is a $m \times n$ matrix where each row corresponds to the projection of all observations in $\bmX$ onto a direction $\bmu_j,j=1,...,m$. 

\medskip

Algorithm~\ref{alg:ParallelDirections} provides a simplified view of data set projection $\bmX\bmU^\top$ as a matrix multiplication, which can be performed in parallel according to hardware constraints \citep{navarro2014survey}. Note that the only non-parallel(izable) loop (over $l$) requires values from previous iteration(s) for the final sum, an operation known as \texttt{reduce}, which is also potentially parallelizable. In what follows, $(\bmU_{j,\cdot})^\top$ (and similar) naturally denote the (column) vector corresponding to pointwise projection of $\bmX$ onto $\bmu_j$, \textit{i.e.}, $j^{th}$ row of matrix $\bmU$.

\medskip

\begin{algorithm}[!h]
\footnotesize
\caption{Parallel projection}
\begin{algorithmic}[0]
\Function{ProjParallel}{$\bmX$, $\bmU$}
\State \textit{\textbf{Initialisation}}
\State  $P \gets \boldsymbol{0}_{m \times n}$
\State \textit{\textbf{Step 1:} Parallel loop through directions}
\For {j $\gets$ 1 \textbf{to} $m$} \textbf{(parallel)}
\State \textit{\textbf{Step 2:} Loop through all points}
\For {i $\gets$ 1 \textbf{to} n} \textbf{(parallel)}
\For {l $\gets$ 1 \textbf{to} d}
\State $\bmP_{j,i}=\bmP_{j,i} + \bmX_{i,l} \cdot \bmU_{j,l}$
\EndFor
\EndFor
\EndFor
\State \Return $\bmP$
\EndFunction
\end{algorithmic}
\label{alg:ParallelDirections}
\end{algorithm}

\begin{algorithm}[!h]
\footnotesize
\caption{Massively-parallel projection}
\begin{algorithmic}[0]
\Function{ProjMassive}{$\bmX$, $\boldsymbol{U}$, $\kappa$}
\State \textit{\textbf{Initialisation}}
\State $\mathcal{B}\gets\varnothing$ \Comment{Queue of threadblocks}
\State  $\boldsymbol{P} \gets \boldsymbol{0}_{m \times n}$
\State Load $\bmX, \boldsymbol{U}$ into the (GPU's) shared memory
\State $t \gets 0$ \Comment{Thread count}
\State $b \gets 0$  \Comment{Threadblock count}
\State \textbf{\textit{Step 1}}: Create threads
\For {$i\gets$ 1 \textbf{to} $n$} \Comment{Loop over the data set}
\For {$j\gets$ 1 \textbf{to} $m$} \Comment{Loop over the directions} 
\State \textbf{Create thread} $\mathbb{t}: \boldsymbol{P}_{j,i} \gets \bmX_{i,\cdot} (\boldsymbol{U}_{j,\cdot})^\top$
\If{$t \text{ mod } \kappa= 0$}
\State $t \gets$ 0 \Comment{Reset counting for another threadblock}
\State $b\gets b+1$ 
\EndIf
\State Add $\mathbb{t}$ to $\mathcal{B}[b]$
\State $t\gets t + 1$
\EndFor
\EndFor
\State \textbf{\textit{Step 2}}: Run threads in parallel (on GPU)
\For{$k\gets$ 1 \textbf{to} b}
\State Load $\mathcal{B}[k]$ into available \emph{streaming multiprocessor}
\State Start executing (in parallel) threads of $\mathcal{B}[k]$
\While{no \emph{streaming multiprocessor} is available} 
\State wait for a free \emph{streaming multiprocessor}
\EndWhile 
\EndFor
\State \Return $\bmP$
\EndFunction
\end{algorithmic}
\label{alg:ParallelGPUMulti}
\end{algorithm}

\begin{figure}[!h]
    \centering
    \includegraphics[trim=0 0 1cm 0,clip=true,width=0.9\linewidth]{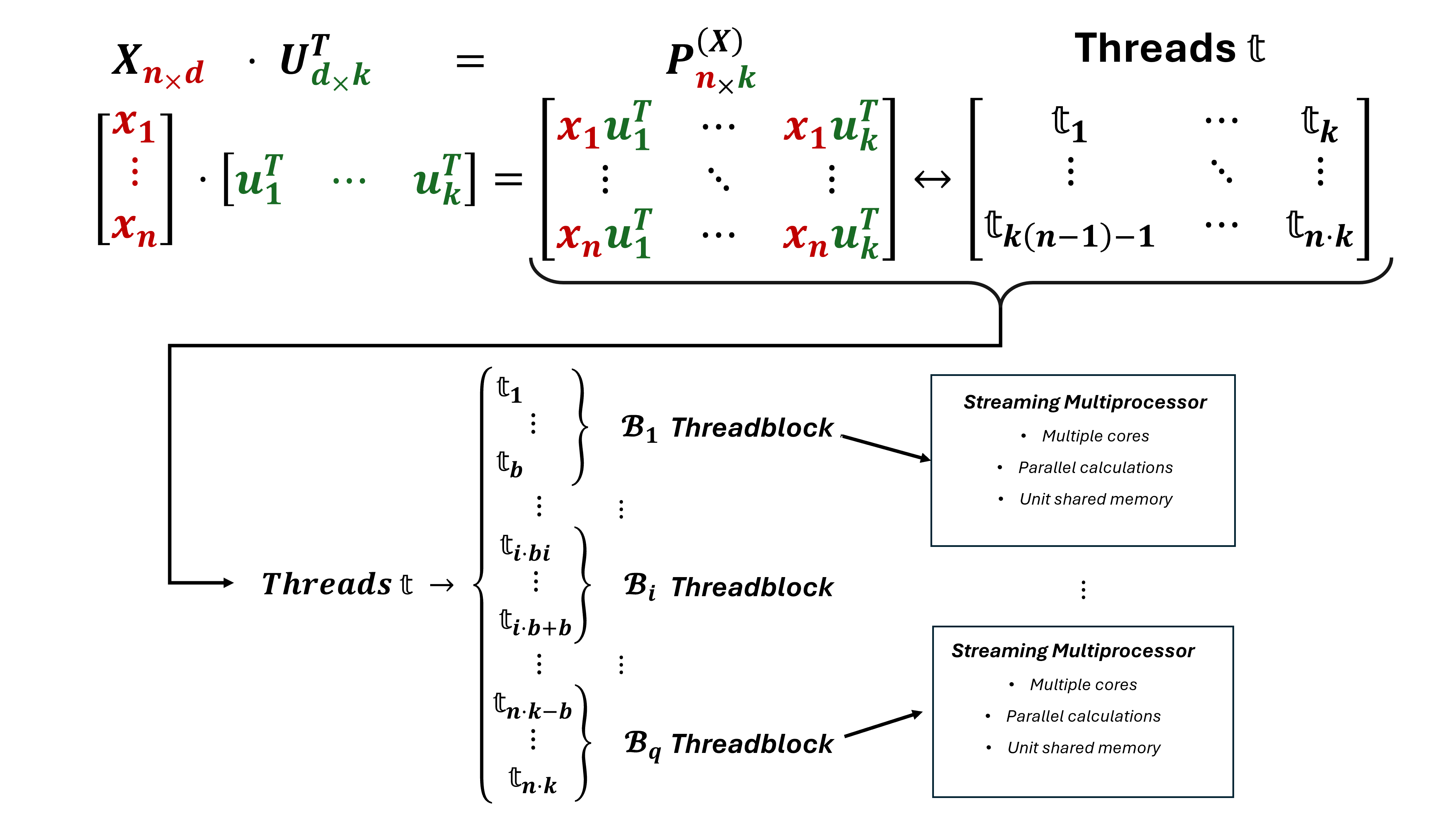}
        \caption{Thread creation and parallel projection of $\bmX\bmU^\top$ on the GPU.}
    \label{fig:BaseProjCUDA}
\end{figure}

While the main focus is on data projection, the operations of direction generation and computation of univariate depth can also be parallelized. With a primary interest in massive parallelization, preferably using GPU architectures, algorithmic tasks should be broken down into simple algebraic operations translatable into GPU-runnable  instructions. The breakdown of the projection task for further processing on a GPU is illustrated in Figure~\ref{fig:BaseProjCUDA}, and its implementation is suggested in Algorithm~\ref{alg:ParallelGPUMulti}. 

\medskip

For relevant details on GPU architecture, the reader is referred to Section \ref{sec:GPU} of the Supplementary Material. Here, we mention several crucial details. Even with powerful infrastructure, resources have hierarchies and limitations. The hierarchy is expressed by threadblocks (assembling multiple threads), and the computational limitation (excluding memory limitations for simplicity) is represented by the maximum number of threadblocks that can run simultaneously within a \emph{streaming multiprocessor} (see Figure~\ref{fig:BaseProjCUDA}).

\subsection{The complete massively-parallel algorithm}\label{ssec:entireAlg}

Algorithm~\ref{alg:ParallelRRS} integrates the previously presented logic into a single algorithm for the massive parallelization of depth computations that satisfy the projection property. The function \texttt{ProjMassive} can be easily applied to a single point, for example, if it is represented as a $1 \times d$ matrix. Not only can the projection operation be parallelized, but the generation of directions (see Section \ref{sec:GPU} in the Supplementary Material for more details and a precise algorithm) and the computation of univariate depths can also be parallelized in a similar manner. Algorithm~\ref{alg:ParallelRRS} is generic with respect to the depth notion, as long as it satisfies the projection property.

\section{Hyperparameter fine-tuning}\label{sec:fineTun}

The second part of our methodology focuses on exploring hyperparameters for RRS in large-scale scenarios, addressing a gap in the literature. Fine-tuning is necessary for two main reasons: to ensure the best convergence of the algorithm to the lowest depth and to optimize the convergence for the shortest runtime. For RRS, we study the optimization based on three hyperparameters: the spherical shrinkage coefficient $\alpha$, the number of refinements $r$, and the total number of used directions $k$. Our goal is to understand how these hyperparameters impact the convergence to the minimum value.

\medskip

In our exploration, a multivariate normal distribution is used for the base and the points analyzed, $\displaystyle X\sim {\mathcal {N}}(\boldsymbol{\mu} ,\boldsymbol{\Sigma})$, centered at the origin ($\boldsymbol{\mu} = 0^d$). The covariance matrix $\boldsymbol{\Sigma}$ is a Toeplitz matrix defined as $\boldsymbol{\Sigma} = (\sigma_{ij})_{\substack{i = 0, \dots, d-1\\j = 0, \dots, d-1}}$, with $\sigma_{ij} = 2^{-|i-j|}$.

\begin{algorithm}[h]
\footnotesize
\caption{Massively-parallel refined random search}
\begin{algorithmic}[0]
\Function{ProjDepthMassive}{$\bmz$,$\bmX$,$k$,$r$,$\alpha$,$\kappa$}
\State \textit{\textbf{Initialisation}}
\State $\bmp\gets\boldsymbol{\mathrm{e}}_1$
\State $m \gets \lceil\frac{k}{r} \rceil$
\State $\epsilon \gets \frac{\pi}{2}$ 
\State $D_{min} \gets$ 1
\For{$l\gets$ 1 \textbf{to} $r$}
\State \textbf{\textit{Generation}}: Generate $m$ random directions ($\epsilon$-around $\bmp$)
\State $\boldsymbol{U} \gets GenMassive(\boldsymbol{p}, m, \epsilon)$  \Comment{Parallel directions' generation}
\State \textbf{\textit{Projection}}: Project $\bmX$ and $\boldsymbol{z}$ onto all directions $\bmU$
\State $\bmP^{(\bmX)} \gets ProjMassive(\bmX,\bmU,\kappa)$
\State $\bmP^{(\bmz)} \gets ProjMassive(\bmz,\bmU,\kappa)$
\State \textbf{\textit{Computation}}: Compute $m$ univariate depths within projections 
\For{$j \gets$ 1 \textbf{to} $m$} \textbf{(parallel)}
\State $D \gets D\bigl((\bmP^{(\bmz)})_{j}|(\bmP^{(\bmX)})_{j,\cdot}\bigr)$ \Comment{Compute a single univariate depth}
\If{$D < D_{min}$}
\State $D_{min} \gets D$ \Comment{Update the depth and its corresponding direction}
\State $\bmp \gets \bmU_{j,\cdot}$
\EndIf
\EndFor
\State $\epsilon \gets \epsilon\cdot\alpha$
\EndFor
\State \Return $D_{min}$
\EndFunction
\end{algorithmic}
\label{alg:ParallelRRS}\end{algorithm}

\subsection{Convergence and spherical cap shrinkage}

First, we study how $\alpha$ affects the final depth value. A data set of 50,000 samples is constructed using a multivariate normal distribution, with the spacial dimension fixed at $d=50$. From this data set, 50 points were randomly selected and analyzed. For each point, a reference value for the (true) minimal depth was computed using 100 refinements, $\alpha$ set to 0.9, and 3,000,000 directions. The minimal value was selected over three such different runs, each with distinct random directions.

\medskip

\begin{figure}[!h]
    \centering
    \includegraphics[width=150mm,scale=0.5]{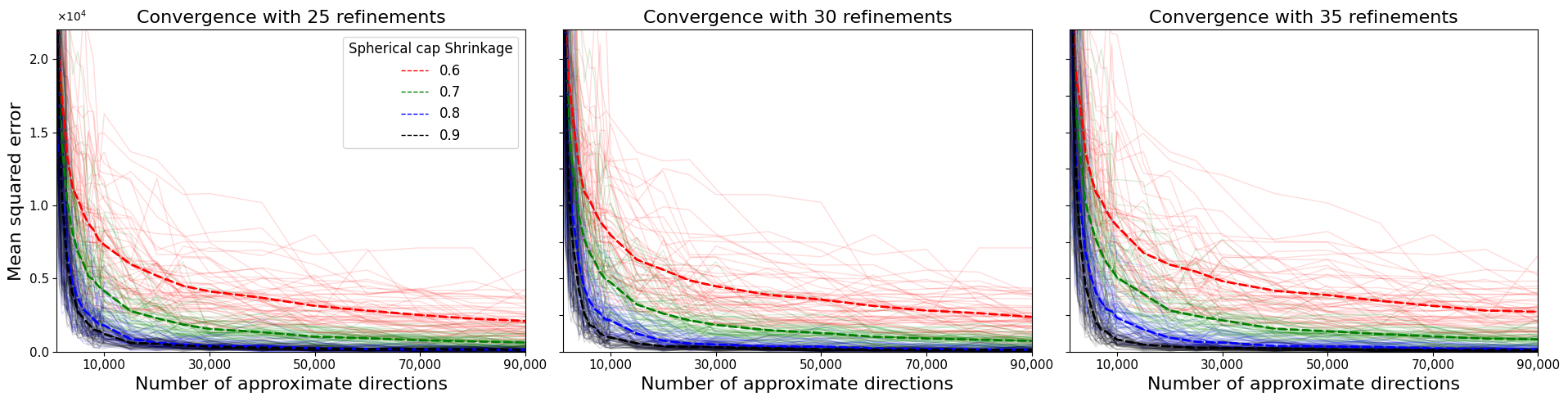}
    \caption{Convergence study using MSE for $D_P$ with respect to a grid of refinements, directions, and spherical shrinkage values, with $d=50$. The number of refinements takes value in \{25, 30, 35\}; the number of directions range from 200 to 90,000 and $\alpha \in\{0.6,0.7,0.8,0.9\}$. The data set size is fixed at 50,000. MSE is computed for 50 random points from $\bmX$ with respect to a reference value. Light and continuous lines represent the MSE of each point, while bold dashed lines stand for the mean value of all MSEs.}
    \label{fig:ConvPrecAlphaRefDir}
\end{figure}

A grid of hyperparameters was used to study the convergence to the minimum value found in the described above reference. Spherical shrinkage values ranged from 0.6 to 0.9, the number of directions ranged from 200 to 90,000, and the number of refinements ranged from 1 to 100. Results for refinements $r\in\{25,30,35\}$ are shown in Figure~\ref{fig:ConvPrecAlphaRefDir}, where the mean squared error (MSE) between the minimal value and computed value is measured  (see Section \ref{Sec:Depth} of the Supplementary Material for further results.)

\medskip

Overall, higher $\alpha$ values indicate faster and better convergence to minimal depth. This pattern is observed for all refinement choices and across all analyzed points. Additionally, the volatility between the mean MSE and individual values decreases with higher $\alpha$ values. A slower decrease in the search surface (corresponding to higher values of $\alpha$) may help to avoid getting trapped into local minima and explore a greater part of $\mathbb{S}^{d-1}$, resulting in better RRS convergence. For the subsequent analysis, $\alpha$ is fixed at 0.9, allowing us to focus on the effects of refinements and directions.

\subsection{Interdependence of refinements and directions}

In this subsection, we explore the impact of refinements and directions on the convergence of the optimization routine. These variables are studied together because they have a more direct correlation with each other. Specifically, $m$, the number of directions used in each refinement, is computed as $\lceil\frac{k}{r}\rceil$. 

\medskip

This investigation is conducted using a grid of values, where the number of directions ranges from 200 to 100,000, the number of refinements ranges from 1 to 175, and the number of dimensions ranges from 5 to 175. The data set $\bmX$ was constructed from the same multivariate normal distribution as previously, containing 10,000 points. From this data set, 50 random points were selected for analysis. For each point in each dimension, the lowest computed depth was taken as its reference value. All computed depths are compared with this reference to calculate the MSE. 

\medskip

\begin{figure}[!h]
    \centering
    \includegraphics[width=150mm,scale=0.5]{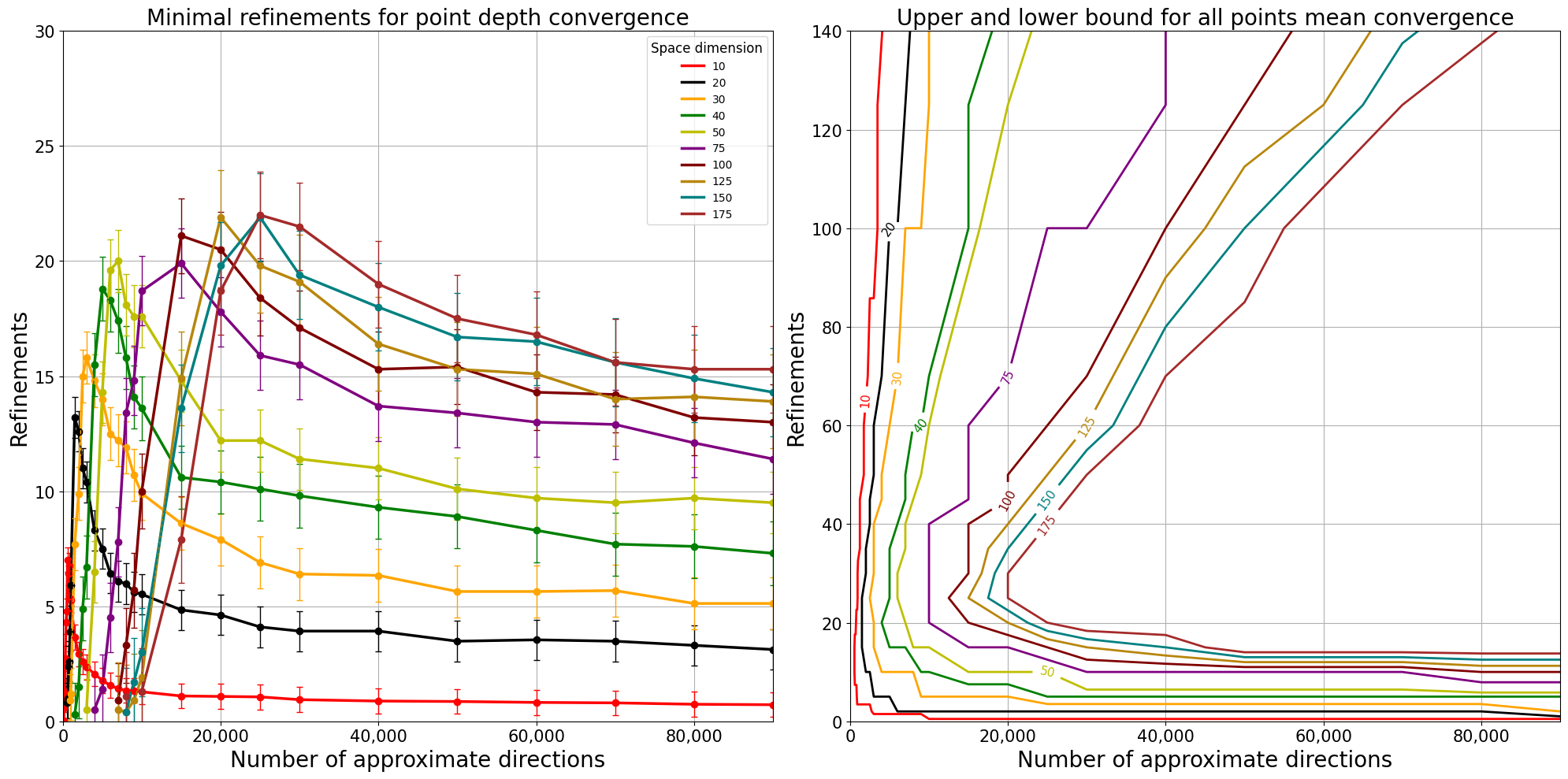}
    \caption{Convergence of the algorithm regarding the interdependence of directions and refinements, with MSE$\leq 10^{-4}$ for a multivariate normal distribution. Depth is computed using projection depth for a data set of 10,000 samples. The number of refinements ranges from 10 to 175, the number of dimensions ranges from 1 to 175, and the number of directions ranges from 200 to 100,000. Projection depth is computed for 50 random points from the distribution. Left: Minimal refinement for point-wise convergence. Right: Upper and lower boundaries for the depth convergence of all points with respect to space dimension.}
    \label{fig:NormConvergenceBounds}
\end{figure}

Figure \ref{fig:NormConvergenceBounds} shows RRS convergence with two distinct views. On the left, curves represent the mean number of refinements with a fixed number of directions for points to converge. Only points with $MSE\leq10^{-4}$ (considered as converged) are taken into account. On the right, curves represent the convergence of all analyzed points. On the left side of each curve, not all points have converged yet, and we take the mean number of minimal refinements for converged points. On the right side, all points have $MSE\leq10^{-4}$, and we analyze the combination of refinements and directions needed for convergence when all points have already converged. 

\medskip

The left graph indicates that not all points converge at the same rate; some require fewer refinements and directions, while others may need a combination of higher refinements and directions. The minimal refinement required reaches a plateau, indicating the convergence of all points, which can be seem in both graphs. The right plot shows that a minimal quantity of directions is needed for all points to converge in every dimension. For the number of refinements with fixed directions, the right plot suggests the existence of upper and lower bounds based on the space dimension. A lower refinement boundary is necessary for the searched region to converge, but a high refinement boundary may indicate an insufficient quantity of directions in each refinement for convergence. 

\medskip

\begin{figure}[!h]
    \centering
    \includegraphics[width=150mm,scale=1]{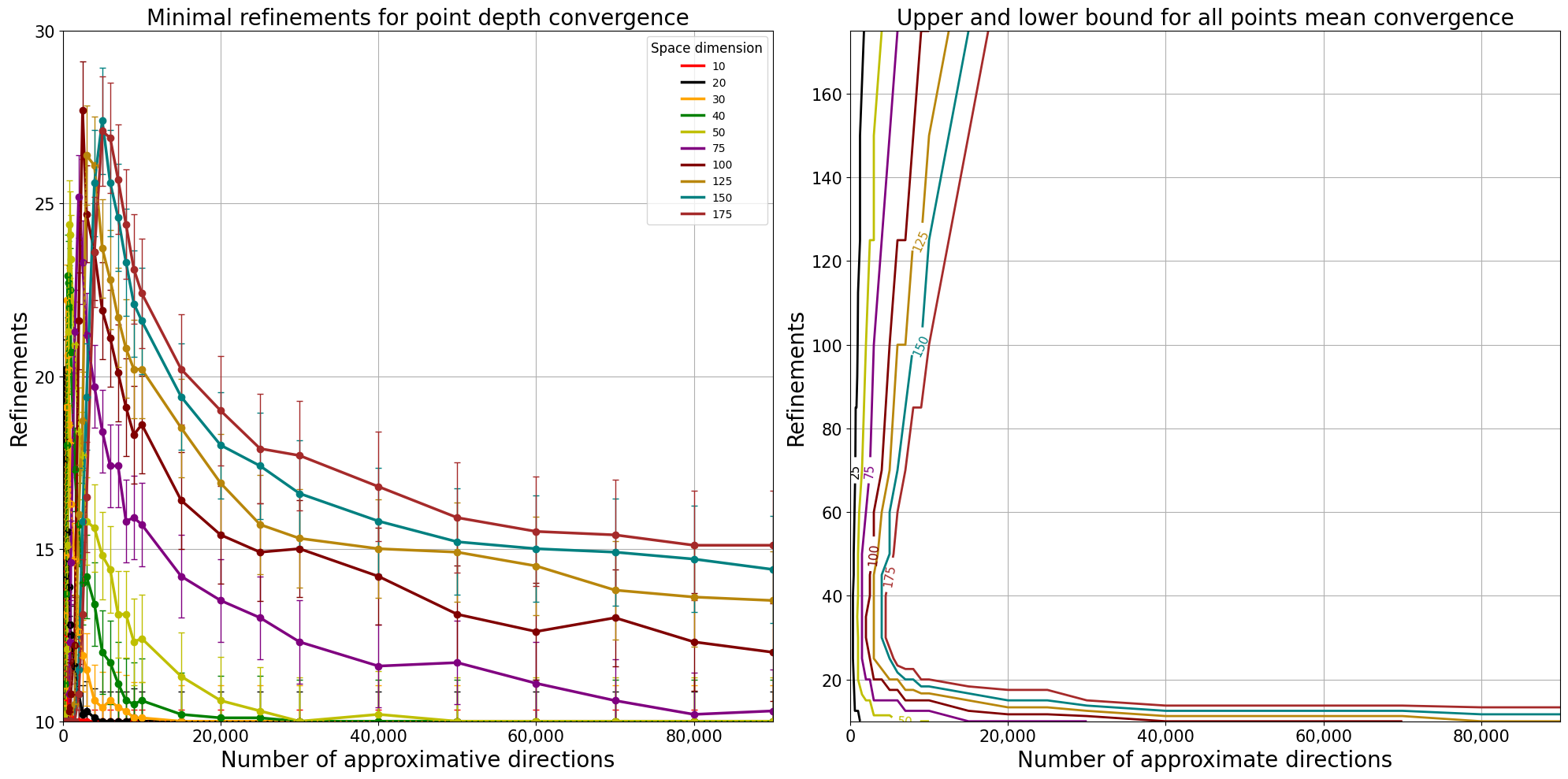}
    \caption{Convergence of the algorithm regarding the interdependence of directions and refinements with MSE $\leq 10^{-4}$ for an exponential distribution. Depth is computed using projection depth for a data set of 10,000 samples. The number of refinements ranges from 10 to 175; the number of dimensions ranges from 1 to 175,and the number of directions ranges from 200 to 100,000. Projection depth is computed for 50 random points from the distribution. Left: Minimal refinement required for point-wise convergence. Right: Upper and lower boundaries for the depth convergence of all points.}
    \label{fig:ExpConvergenceBounds}
\end{figure}

We also explore convergence regions for a multivariate exponential distribution (Figure~\ref{fig:ExpConvergenceBounds}). For this distribution, it seems that fewer directions and refinements are needed to reach the lowest depth. Convergence frontiers are closer together on the right side of the figure, while the left part shows earlier convergence even for higher dimensions. 
 
\medskip

This methodology for hyperparameter selection helps us to understand how the algorithm behaves in higher dimensions and determine the optimal parameter combinations for depth convergence. Our next goal is to verify time gains by comparing both approaches and further developing~\eqref{eqn:Sp} to explain the observed time behaviors.

\section{Runtime analysis in large-scale scenarios}
\label{sec:timeStudy}

In this section, we investigate the runtime performance in large-scale scenarios and evaluate how the observed temporal behavior aligns with our proposed equations.

\subsection{Breakdown time analysis}

We analyze the significance of each operation highlighted in~\eqref{eqn:GPU_time} by measuring their individual runtime contributions. This analysis employs three different approaches to examine the relative impact of each operation on the total runtime.

\medskip

\begin{figure}[!h]
    \centering
    \includegraphics[width=150mm,scale=0.5]{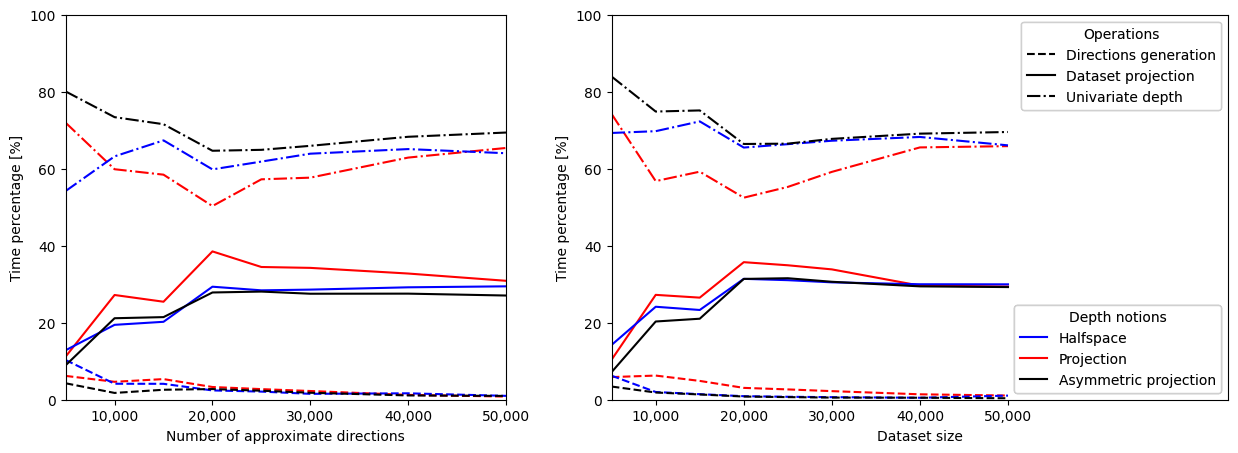}
    \caption{Breakdown time percentage for projection-based depth notions with refinements fixed at 1 and dimensions at 150. The red line represents $D_P$, blue lines $D_H$, and black lines $D_{AP}$. Dashed lines indicate direction generation time, solid lines represent data set projection $\bmX$, and dot-dashed lines denote univariate depth computation. Left: $n$ is fixed at 10,000 while the number of directions is varied. Right: The number of directions $k$ is fixed at 10,000 while the data set size is varied.}
    \label{fig:BdTimeDirBase}
\end{figure}

Figure~\ref{fig:BdTimeDirBase} illustrates the first two approaches, where the time breakdown for the three depth notions ($D_P$, $D_H$, and $D_{AP}$) is presented. The results are averaged over 100 points. This setup allows us to study the impact of varying the number of approximate directions. Overall, univariate depth computation emerges as the rate-determining step. This contrasts with the standard RRS results shown in Figure \ref{fig:BdTimeCpp}, where projecting the data set consumed a higher percentage of the total time.

\medskip

Projecting the data set is a significant factor in the overall runtime, accounting for approximately one-third of the total time. This demonstrates how our algorithm has decreased the projection time to the point where it is no longer the rate-determining operation. This change is evident in the inversion of the impacts between univariate depth computation and data set projection. 

\medskip

Interestingly, even with $d=150$, the number of dimensions does not significantly affect the final runtime, as the univariate computation dominates the depth calculation time. Parallelizing these operations results in a final time decomposition that differs from traditional time complexity analysis, where we observe that $Time(\mathcal{O}(knd))<Time(\mathcal{O}(kn))$. These findings suggest that the cost of projecting the data, $C_P$, is lower than the cost of univariate depth, $C_{d(1)}$ (since matrix multiplication, which consists of FLOPS, can be more efficiently parallelized on the GPU). Returning to our model, when $d$ is fixed, both data projection and univariate depth computation scale with $n\cdot k$, exhibiting similar performance trends as either $k$ or $n$ increases.

\medskip

\begin{figure}[!h]
    \centering
    \includegraphics[width=100mm,scale=0.5]{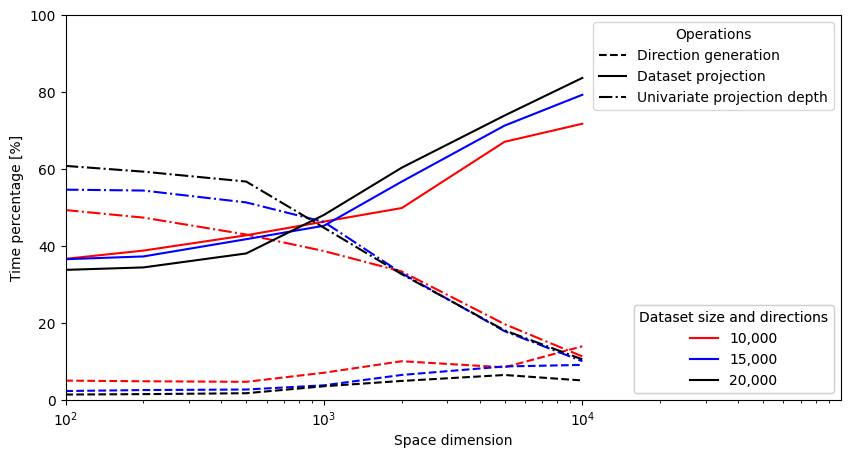}
    \caption{Breakdown of operation times for projection depth with respect to space dimension. The number of directions and data set size are fixed according to line color: red for $n,k=10,000$, blue for $n,k=15,000$, black for $n,k=20,000$. Operations are indicated by line style: dashed lines represent direction generation, solid lines represent data set projection, and dot-dashed lines indicate univariate depth computation.}
    \label{fig:BdTimeDimension}
\end{figure}

Finally, space dimension is also crucial for time computation and complexity. Figure~\ref{fig:BdTimeDimension} provides a breakdown of the main steps for $D_P$ calculation. This computation is performed with one refinement, where the number of approximate directions and the data set size are fixed at equal values. As $d$ increases, measures based on projected values become less important, while data set projection regains its role as the most significant operation for depth approximation in higher dimensions. The importance of generating directions also increases with higher dimensions, surpassing the runtime impact of univariate depth computation. However, it remains overshadowed by the data set projection operation. 

\medskip

These results are key for understanding the importance of each factor in Equation \ref{eqn:GPU_time}, and how the algorithm performs in extreme environments. Generating directions becomes more significant in high-dimensional cases but is overshadowed by other operations. Univariate depth computation emerges as the primary time-consuming operation overall, losing its dominance to data set projection only in very high dimensions (around $10^3$ variables). Given a consistent computer configuration, the algorithm behavior aligns with the equations proposed in Section~\ref{sec:extPara}. 

\medskip

Beyond the breakdown time analysis, the proposed equations describing the speedup factor~\eqref{eqn:Sp} can also be validated with empirical results.

\subsection{Empirical speedup analysis} 

In this part, we compare the runtime of both approaches using the same hyperparameters, and we verify how well the empirical results match the theoretical speedup factor described in~\eqref{eqn:Sp}.

\medskip

\begin{figure}[!h]
    \centering
    \includegraphics[width=150mm,scale=0.5]{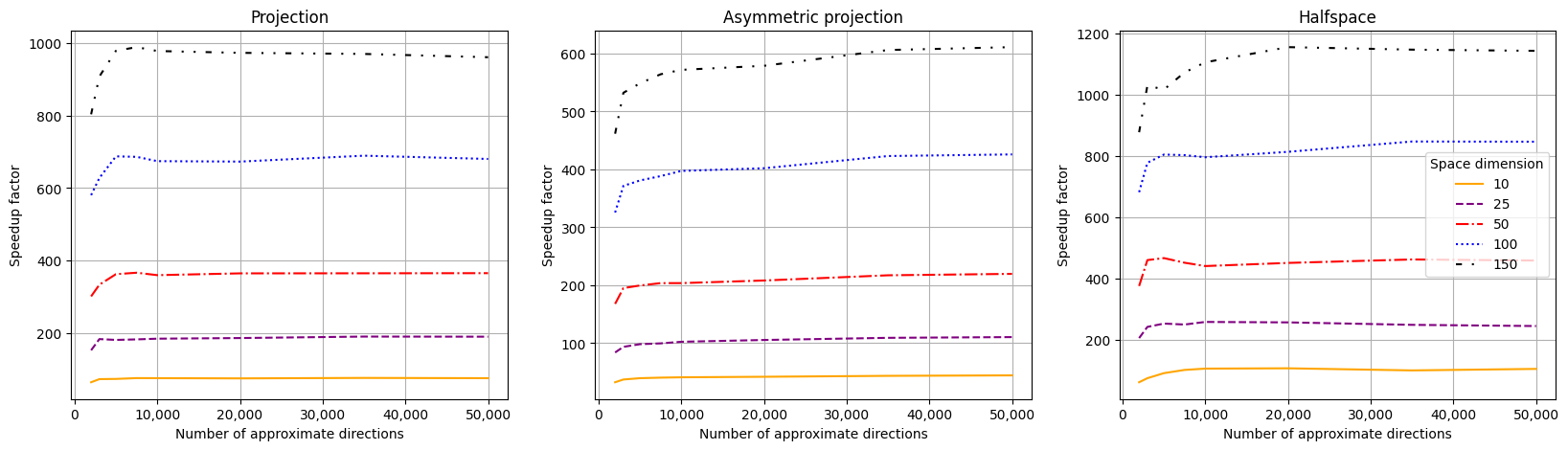}
    \caption{Speedup factor for projection-based depth notions as a function of approximating directions and space dimension. From left to right: halfspace, projection, and asymmetric projection depth. Ordinate scale are set to better accommodate computed values providing better visualization. Colors represent the space dimension, orange $d=10$, purple $d=25$, red $d=50$, blue $d=100$, and black $d=150$.}
    \label{fig:speedupNotions10k} 
\end{figure}

Different notions of depth will exhibit varying speedup factors, as they are influenced by $\mathcal{D}$, as formalized in~\eqref{eqn:CPU_time} and~\eqref{eqn:GPU_time}. The variation in speedup for the three notions is illustrated in Figure~\ref{fig:speedupNotions10k}.
Overall, the speedup factor increases with the number of directions until it reaches a plateau, which depends on the dimension $d$ and the depth notion. This result is aligned with~\eqref{eqn:Sp}. Note that the speedup scale in the graphs is adjusted for better  visualization of the phenomenon.  

Using previous results, we can develop $S_p$. Given that $C_{rv}$ and $C_c$ are negligible compared to other operations and considering that $\mathcal{D}\sim m\cdot n$, the speedup factor becomes: \[ S_p=\frac{m\cdot n\cdot\left(C_p\cdot d+C_{d(1)}\right)}{\lceil\frac{m\cdot n}{g}\rceil \cdot \lambda \cdot\left(C_p\cdot \lceil\frac{d}{d_{max(m,n)}}\rceil+C_{d(1)}\right)}.\] 

\medskip

From this equation, we observe two key behaviors. First, the increase in $S_p$ occurs for smaller values of $m\cdot n$. As this product increases, the function shows an amortized convergence to the maximum speedup value, with small fluctuations perceived as a plateau value. The latter depends only on the dimension $d$. To better illustrate this plateau, we consider $m\cdot n$ as a multiple of $g$, \textit{i.e.},  $m\cdot n=w \cdot g$, where $w$ is a factor that cancels out, resulting in \eqref{eqn:SpeedupSimDim}:

\begin{equation}
\label{eqn:SpeedupSimDim}
S_p=\frac{g}{\lambda}\cdot\frac{C_p\cdot d+C_{d(1)}}{C_p\cdot\lceil\frac{d}{d_{max(m,n)}}\rceil+C_{d(1)}}.
\end{equation}

\begin{figure}[!h]
    \centering
    \includegraphics[width=150mm,scale=0.5]{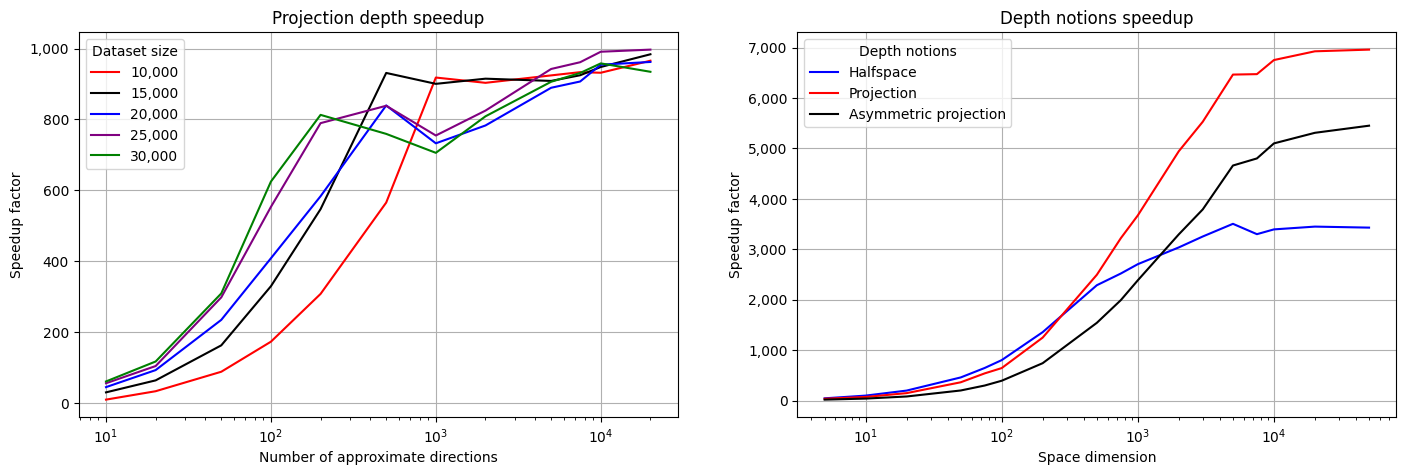}
    \caption{Speedup factor analysis. Left: Projection depth with increasing approximate directions and data set size, with dimension fixed at 150; each color represents a different data set size. Right: Speedup factor for different projection-based notions as the dimension  increases. Both directions and data set samples are fixed at 10,000. In both cases, computations are performed with one refinement.}
    \label{fig:speedupDpDirBase}
\end{figure}

As stated, this behavior depends on the product $m\cdot n$ and the dimension, which is supported by the results shown in the left graph of Figure~\ref{fig:speedupDpDirBase}, exploring $D_P$ computation. With $d$ fixed at $150$, approximately the same $S_p$ plateau value is reached by increasing either $n$ or $m$. The rate at which the plateau is reached depends on $n\cdot m$, so the curves do not overlap with lower numbers of directions. Time was measured as the mean time over 100 points. 

\medskip

Continuing the exploration of projection notions, the impact of $d$ on $S_p$ is examined in the right graph of Figure~\ref{fig:speedupDpDirBase}. Time is computed with a fixed data set of 10.000 samples, 10.000 directions, and one refinement. The space dimension is increased from 5 to 50.000 in smaller increments than in previous explorations. A similar pattern is observed, showing an increase in speedup up to a point, followed by a fluctuating plateau where the speedup value oscillates around this value with an amortized convergence. This behavior is similar to how $\frac{d}{\lceil\frac{d}{d_{max(m,n)}} \rceil}$ behaves like $\frac{m\cdot n}{\lceil\frac{m\cdot n}{g}\rceil}$ in~\eqref{eqn:SpeedupSimDim}.

Compared to previous state-of-the-art approaches, we achieved a time reduction of three orders of magnitude, with the importance of data set projection and univariate depth inverted. In the next section, we shift our focus from comparing $T_C$ with $T_G$ to analyzing and visualizing the absolute runtime values observed with different parameters.

\subsection{Numerical time analysis}

In these experiments, computational time is measured for a grid of parameters including the number of directions, dimensions, data set sizes, and refinements. These measurements are conducted using $D_P$, since all depth notions exhibit similar time behavior. The results (in seconds) are presented in Table~\ref{tab:timeTableProj10k}, where $D_P$ is evaluated with a data set of 10,000 samples and a single refinement.

\medskip

The time appears to grow proportionally with the number of directions after a threshold of approximately 5,000 directions, aligning with the observations in Figure \ref{fig:speedupNotions10k}. Interestingly, the dimension seems to have little to no impact on the final runtime, except when the number of approximate directions is high. The last row deviates from this proportional pattern, exhibiting a higher runtime than expected. 

\noindent
\begin{table}[!h]
\footnotesize
    \centering
    \begin{tabular}{ccccccccc}
    &\multicolumn{7}{c}{\textbf{\textit{Space dimension}}}\\
    \cline{2-9}\\
\multicolumn{1}{c|}{\textit{\textbf{Directions}}} & \textbf{5}  &  \textbf{10}  &  \textbf{25}  &  \textbf{50}  &  \textbf{75}  &  \textbf{100} & \textbf{125}  &  \textbf{150} \\
\hline
\multicolumn{1}{c|}{\textbf{\textit{1,000}}} & 
0.01 & 0.01 & 0.01 & 0.01 & 0.01 & 0.01 & 0.01 & 0.01 \\
\multicolumn{1}{c|}{\textbf{\textit{2,500}}} & 
0.01 & 0.01 & 0.01 & 0.01 & 0.01 & 0.01 & 0.01 & 0.01 \\
\multicolumn{1}{c|}{\textbf{\textit{5,000}}} & 
0.02 & 0.02 & 0.02 & 0.02 & 0.02 & 0.02 & 0.02 & 0.02 \\
\multicolumn{1}{c|}{\textbf{\textit{10,000}}} &   
0.03 & 0.03 & 0.03 & 0.03 & 0.03 & 0.03 & 0.03 & 0.03 \\
\multicolumn{1}{c|}{\textbf{\textit{25,000}}} &   
0.06 & 0.07 & 0.07 & 0.07 & 0.07 & 0.07 & 0.07 & 0.08 \\
\multicolumn{1}{c|}{\textbf{\textit{50,000}}} &   
0.13 & 0.14 & 0.14 & 0.14 & 0.14 & 0.15 & 0.15 & 0.16 \\
\multicolumn{1}{c|}{\textbf{\textit{100,000}}} &  
0.26 & 0.28 & 0.28 & 0.30 & 0.28 & 0.28 & 0.30 & 0.30 \\ 
\multicolumn{1}{c|}{\textbf{\textit{250,000}}} &  
\textbf{2.49} & \textbf{2.49} & \textbf{2.52} & \textbf{2.59} & \textbf{2.60} & \textbf{2.60} & \textbf{2.64} & \textbf{2.66} \\
    \end{tabular}
        \caption{Computational time (in seconds) to calculate depth for a data set of 10,000 points using one refinement.}
    \label{tab:timeTableProj10k}
\end{table}

This time increase can be partially explained by referring back to~\eqref{eqn:SpeedupSimDim}. After reaching the maximum speedup, the final change in value $S_p$ is $\lceil \frac{d}{d_{max(m,n)}} \rceil$, indicating memory saturation. However, the most significant increment is in the univariate depth computation due to its memory-bound operations, flow divergence, and low-level conditional steps \citep{fung2007dynamic}. Data set projection decreases in importance for the final runtime, dropping from approximately 30\% to around 16\%, while univariate depth computation increases from 70\% to 83\% of the final time. 

\medskip

Refinements can play a crucial role in computational time, as they not only introduce the optimization character into the algorithm, but can also reduce the memory used in each refinement, mitigating memory overload and potential efficiency loss. Table~\ref{tab:RefinementTime} shows the computational time for $D_P$ with a 10,000-point data set using 250,000 directions, with the number of refinements ranging from 1 to 100. Significant time reductions are observed with the inclusion of more refinements, which helps alleviate memory overload and brings the runtime closer to the expected values from Table~\ref{tab:timeTableProj10k}. Beyond the initial decrease, further increases in refinements do not significantly affect the runtime.

\begin{table}[!h]
\footnotesize
    \centering
    \begin{tabular}{lcccccccc}
    &\multicolumn{7}{c}{\textbf{\textit{Space dimension}}}\\
    \cline{2-9}\\
\multicolumn{1}{c|}{\textit{\textbf{Refinements}}} & 5  &  10  &  25  &  50 & 75 & 100 & 125 & 150 \\
\hline
\multicolumn{1}{c|}{\textbf{\textit{1}}} &  
2.49 & 2.49 & 2.52 & 2.59 & 2.60 & 2.60 & 2.64 & 2.66 \\
\multicolumn{1}{c|}{\textbf{\textit{10}}} &  
0.62 & 0.67 & 0.67 & 0.69 & 0.69 & 0.69 & 0.74 & 0.76 \\
\multicolumn{1}{c|}{\textbf{\textit{25}}} &  
0.61 & 0.65 & 0.65 & 0.68 & 0.69 & 0.69 & 0.70 & 0.71 \\
\multicolumn{1}{c|}{\textbf{\textit{50}}} &  
0.61 & 0.65 & 0.65 & 0.67 & 0.68 & 0.68 & 0.70 & 0.73 \\
\multicolumn{1}{c|}{\textbf{\textit{100}}} &  
0.66 & 0.67 & 0.67 & 0.68 & 0.70 & 0.69 & 0.71 & 0.72 \\
    \end{tabular}
        \caption{Computational time (in seconds) to calculate $D_P$ with a data set of 10,000 samples and 250,000 approximate directions.}
    \label{tab:RefinementTime}
\end{table}

The size of the data set is also a key factor in computational time. Table~\ref{tab:BaseSizeTime} focuses on varying the data set size while keeping refinements fixed at 1 and the number of directions fixed at 10,000. Our results indicate a similar behavior between increasing the number of directions and increasing the data set size, as expected from~\eqref{eqn:GPU_time}.

\begin{table}[!h]
\footnotesize
    \centering
    \begin{tabular}{lcccccccc}
    &\multicolumn{7}{c}{\textbf{\textit{Space dimension}}}\\
    \cline{2-9}\\
\multicolumn{1}{c|}{\textit{\textbf{Data set size}}} & 5  &  10  &  25  &  50 & 75  & 100 & 125  &  150\\
\hline
\multicolumn{1}{c|}{\textbf{\textit{1,000}}}& 
0.01 & 0.01 & 0.01 & 0.01 & 0.01 & 0.01 & 0.01 & 0.01 \\ 
\multicolumn{1}{c|}{\textbf{\textit{2,500}}} &
0.01 & 0.01 & 0.01 & 0.01 & 0.01 & 0.01 & 0.01 & 0.01 \\ 
\multicolumn{1}{c|}{\textbf{\textit{5,000}}} &
0.01 & 0.02 & 0.01 & 0.01 & 0.02 & 0.02 & 0.02 & 0.02 \\ 
\multicolumn{1}{c|}{\textbf{\textit{10,000}}} &
0.03 & 0.03 & 0.03 & 0.03 & 0.03 & 0.03 & 0.03 & 0.03 \\ 
\multicolumn{1}{c|}{\textbf{\textit{25,000}}} &
0.06 & 0.06 & 0.07 & 0.07 & 0.07 & 0.07 & 0.08 & 0.08 \\ 
\multicolumn{1}{c|}{\textbf{\textit{50,000}}} &
0.14 & 0.14 & 0.15 & 0.14 & 0.15 & 0.15 & 0.17 & 0.16 \\ 
\multicolumn{1}{c|}{\textbf{\textit{100,000}}} &
0.29 & 0.28 & 0.30 & 0.29 & 0.29 & 0.33 & 0.31 & 0.33 \\ 
    \end{tabular}
        \caption{Computational time (in seconds) to calculate $D_P$ for a data set with 20,000 directions and one refinement.}
    \label{tab:BaseSizeTime}
\end{table}

Overall, the numerical time analysis reveals that the number of directions and data set size have a more significant impact on computational time than the space dimension. Time does not increase significantly with higher dimensions. 

\medskip

Beyond computational time, we also investigate data ranking and ordering with respect to the distribution center to verify the precision of data ordering.

\section{Center-outward ordering precision and depth benchmark}
\label{sec:rankOrder}

The natural ordering provided by depth notions can be evaluated in higher dimensions using ranking correlation techniques. A data set with 100,000 samples is constructed using a normal distribution and a Toeplitz matrix, as previously formalized. Since the distribution is known, its probability density function (pdf) can serve as a reference to ensure the correct ordering of the data. 

\medskip

Three depth notions are used for this ranking assessment: $D_P$, $D_{AP}$, and Mahalanobis depth ($D_M$). The Mahalanobis depth is based on the Mahalanobis distance~\citep{mahalanobis2018generalized} and is computed as: 
$$D_M(\boldsymbol{z}|\boldsymbol{X})=\left(1+(\boldsymbol{z}-\boldsymbol{\mu})'(\boldsymbol{\Sigma}^{-1})(\boldsymbol{z}-\boldsymbol{\mu})\right)^{-1},$$

\noindent where $\boldsymbol{\mu}$ is the mean of $\boldsymbol{X}$ and $\boldsymbol{\Sigma}$ is its variance-covariance matrix.

\medskip

Table~\ref{tab:rankCor} presents the ranking correlation of 5,000 randomly chosen points from the data set. Each point was analyzed using $D_P$ and $D_{AP}$, both using 40 refinements, a total of 100,000 directions, and $\alpha$ set to 0.9 based on the previous analysis. $D_M$ was computed using two different estimators for $\mu$ and $\boldsymbol{\Sigma}$:
\begin{itemize}
    \item $D_{M(L)}$: Uses the maximum likelihood estimator (MLE) for $\boldsymbol{\Sigma}$ and an unbiased estimator for the location, $\mu = \frac{1}{n} \sum_{i=1}^{n} X_i$.
    \item $D_{M(R)}$: Employs a robust estimation method, the minimum covariance determinant (MCD) with a factor of 0.5, for both the location $\mu$ and the scatter $\boldsymbol{\Sigma}$~\citep{rousseeuw2003robust,rousseeuw1999fast}.
\end{itemize}

The study uses two robust ranking correlation techniques, namely Kendall's $\tau$ and Spearman's $\rho$ correlation \citep{SpearmanCorr1987, KendallsTau1938,croux2010influence}. Both $D_P$ and $D_{AP}$ show very strong correlations with the real pdf ranking across all studied dimensions, although the correlation values (as expected) decrease with higher space dimension. Overall, $D_M$ exhibits high correlation values for both estimation approaches, since both techniques are well-suited for normal distributions.
 
\begin{center}
\footnotesize
\begin{tabular}{lcccccc}
\toprule
 &&\multicolumn{5}{c}{\textbf{\textit{Dimensions}}} 
\\\cmidrule(r){1-1}
\cmidrule(r){2-7}
\multicolumn{1}{c}{Correlation} & 5&10&25&50&100&150\\
\bottomrule
\textbf{\textit{Spearman $\rho$ }} \\
PDF x $D_P$ & .9999 & .9998 & .9995 & .9987 & .9963 & .9912 \\
PDF x $D_{AP}$ & .9987 & .9969 & .9903 & .9812 & .9549 & .9140 \\
PDF x $D_{M(L)}$ & .9999 & .9998 & .9998 & .9996 & .9993 & .9990 \\
PDF x $D_{M(R)}$ & .9999 & .9998 & .9998 & .9996 & .9994 & .9990\\
$D_P$ x $D_{AP}$ & .9987 & .9971 & .9907 & .9810 & .9536 & .9115\\ 
\midrule
\textbf{\textit{Kendall's $\tau$}} \\
PDF x $D_P$ & .9929 & .9891 & .9821 & .9687 & .9483 & .9216 \\
PDF x $D_{AP}$ & .9697 & .9530 & .9172 & .8820 & .8157 & .7468 \\
PDF x $D_{M(L)}$ & .9945 & .9921 & .9892 & .9848 & .9793 & .9741 \\ 
PDF x $D_{M(R)}$ & .9962& .9932& .9902 & .9851 & .9802 & .9745 \\
$D_P$ x $D_{AP}$ & .9704 & .9545 & .9187 & .8813 & .8123 & .7431\\

\bottomrule
\end{tabular}
\captionof{table}{Ranking correlation for depth ordering with multivariate Gaussian distribution.}
\label{tab:rankCor}
\end{center}

\medskip

Using the same number of points, directions, $\Sigma$, $\mu$, and $d=50$, we also ranked four Student's $t$ distributions with degrees of freedom $\nu=1, 2, 5, 10$. The results are displayed in Table~\ref{tab:StudentTCor}. For smaller degrees of freedom, $D_P$ and $D_{AP}$ exhibit higher $\rho$ and $\tau$ values than $D_{M(L)}$. Overall, the correlation results are higher with a Student's $t$ than with an underlying Gaussian distribution.

\begin{center}
\footnotesize
\begin{tabular}{lccccc}
\toprule
 &&\multicolumn{2}{c}{\textbf{\textit{$\nu$}}} 
\\\cmidrule(r){1-1}
\cmidrule(r){2-5}
\multicolumn{1}{c}{Correlation}& 1 & 2 & 5 & 10 & Gaussian
\\
\bottomrule
\textbf{\textit{Spearman $\rho$ }} \\
PDF x $D_P$& .9999 & .9998 & .9995 & .9991 & .9987\\
PDF x $D_{AP}$& .9996 & .9992 & .9982 & .9967 & .9812\\
PDF x $D_{M(R)}$ & .9999 & .9999 & .9999 & .9999 & .9996\\
PDF x $D_{M(L)}$ &.9936 & .9978 & .9999 & .9999 & .9996\\
$D_P$ x $D_{AP}$&.9995 & .9990 & .9978 & .9960 & .9810\\

\midrule
\textbf{\textit{Kendall's $\tau$}} \\
PDF x $D_P$& .9907 & .9871 & .9804 & .9741 & .9699\\
PDF x $D_{AP}$& .9836 & .9771 & .9638 & .9514 & .8820\\
PDF x $D_{M(R)}$ & .9973 & .9964 & .9944 & .9928 & .9848\\
PDF x $D_{M(L)}$ & .9332 & .9608 & .9936 & .9934 & .9851\\
$D_P$ x $D_{AP}$& .9817 & .9742 & .9597 & .9461 & .8813\\
\bottomrule
\end{tabular}
\captionof{table}{Ranking correlation for depth ordering with Student's t distribution.}
\label{tab:StudentTCor}
\end{center}

\medskip

Ranking correlation is performed to assess how well the collective centrality of points is measured with respect to their underlying empirical distribution and how these values can be properly ranked. A key factor for optimized results is the choice of hyperparameters, which is based on our exploration method. $D_P$ and $D_{AP}$ can be further fine-tuned to achieve more precise results by reducing the MSE value of the depth computation.

\section{Conclusion}
\label{sec:Conclusion}

This article introduced a novel methodology for the massive parallelization of projection-based depths, addressing the longstanding computational challenges associated with data depth in high-dimensional spaces. By leveraging the Refined Random Search algorithm and modern GPU architectures, our approach achieves unprecedented speedup (up to 7,000 times faster than in the state-of-the-art work by~\cite{dyckerhoff2021approximate}, and this on a laptop graphics card) while maintaining the precision of depth calculations. This advancement enables the practical application of data depth in large-scale and high-dimensional data sets, opening new avenues for robust statistical analysis in fields such as anomaly detection, classification, and dimension reduction. 

\medskip

Our framework decomposes the computation of projection-based depths into parallelizable tasks, focusing on the three main operations: direction generation, data projection, and univariate depth computation. By optimizing these operations for GPU execution, we significantly reduce runtime without compromising accuracy. The empirical results, validated through synthetic data, demonstrate the effectiveness of our approach in handling extreme data scenarios, achieving both efficiency and precision.

\medskip

Furthermore, our study provides insights into hyperparameter fine-tuning, such as the selection of spherical cap shrinkage, the number of refinements, and directions, to optimize convergence and runtime. These findings ensure the robustness and adaptability of our method across different data sets and space dimensions. The center-outward ordering precision, evaluated through ranking correlation techniques, confirms the reliability of our approach in high-dimensional data analysis.

\medskip

The implications of this work are far-reaching. By overcoming the computational bottlenecks of data depth, our methodology paves the way for broader adoption (\textit{e.g.}, also in multivariate functional spaces~\cite{claeskens2014multivariate} or the space of curves~\cite{de2021depth}) of this powerful statistical tool in large-scale applications. Future research could explore the extension of this framework to other depth notions, further optimization of GPU utilization, and the application of our method to real-world data sets. Additionally, investigating the integration of our approach with other statistical techniques, such as robust estimation and machine learning algorithms, could yield even more powerful analytical tools.

\medskip

Finally, we have made the RRS algorithm (and other depth functions) available in the \texttt{\textbf{Python}}\textbf{\textit{-library}} {\bf \href{https://data-depth.github.io/}{data-depth}}, providing researchers and practitioners with the ready-to-use tools to implement and to build upon our work.

\section*{Acknowledgments}
{This research has been conducted within the Research Chair "Digital Finance" under the aegis of the Risk Foundation, a joint initiative by Cartes Bancaires CB, Telecom Paris and University of Paris 2 Panthéon-Assas. The authors further gratefully acknowledge the support of the Young Researcher Grant of the French National Agency for Research (ANR JCJC 2021) in category Artificial Intelligence registered under the number ANR-21-CE23-0029-01 and the support of the CIFRE grant number 2021/1739. We would like to thank seminar participants at the 2024 International Conference on Statistics and Data Science (ICSDS). We also thank Samuel Willy and Hafid Chakir for their support.}

\section*{Supplementary material}
\label{sec:SupMat}

This article contains supplementary materials, which provide:
\begin{itemize}
    \item Additional information about simple random search and refined random search.
    \item Further results and illustrations regarding depth convergence and depth outcomes.
    \item A more detailed explanation of the parallel construction of threads and their key constituents.
\end{itemize}

\bibliographystyle{Chicago}

\newpage
\renewcommand{\thefigure}{S\arabic{figure}}
\renewcommand{\thealgorithm}{S\arabic{algorithm}}
\setcounter{figure}{0}
\setcounter{algorithm}{0}
\appendix

\begin{center}

\LARGE {Supplementary material to the article \\ "Massive parallelization of projection-based depths"}

\large{Leonardo Leone, Pavlo Mozharovskyi, David Bounie}
\end{center}

This supplementary material provides additional information about simple random search and refined random search in Section \ref{Sec:RandSearch}. Section \ref{sec:Depth} offers further results and illustrations regarding depth convergence and depth outcomes. Finally, a more detailed explanation of the parallel construction of threads and their key constituents is presented in Section \ref{sec:GPU}.

\section{Simple random search and refined random search}
\label{Sec:RandSearch}

\medskip
\paragraph{Simple random search.}
\medskip

The first structure for the proposed method is developed using RS. Depth is computed with respect to a set of randomly generated directions using an algorithm called (Simple) Random Search (RS). In RS, directions are randomly generated on the unit sphere ($\mathbb{S}^{d-1}=\{\boldsymbol{x}\in\mathbb{R}^d\,:\,\|\boldsymbol{x}\|=1\}$) following a normal distribution, as shown in Algorithm \ref{alg:randomSphere}. The data set $\boldsymbol{X}$ is then projected, and the respective depth is computed. The smallest depth value for $k$ directions is kept as the approximated depth. 

\begin{algorithm}[!h]
\footnotesize
\caption{Create a random direction on the unit sphere from a normal distribution}
\begin{algorithmic}[0]
\Function{randomSphere}{d}
\State $\boldsymbol{u} \gets \boldsymbol{0}_{d}$  
\State $s\gets 0$ 
\For {i $\gets$ 1 \textbf{to} d}
\State $\boldsymbol{u}_i \gets$ $\mathcal{N}$(0,1) 
\State $s\gets s + \boldsymbol{u}_i^2$
\EndFor
\For{i $\gets$ 1 \textbf{to} d}
\State $\boldsymbol{u}_i\gets\frac{\boldsymbol{u}_i}{\sqrt{s}}$
\EndFor
\State \Return $\boldsymbol{u}_i$
\EndFunction
\end{algorithmic}
\label{alg:randomSphere}
\end{algorithm} 

Algorithm \ref{alg:simpleRandomSearch} presents the respective pseudocode for a generic depth notion computation using RS.

\bigskip

\begin{algorithm}[!h]
\footnotesize
\caption{Computing depth for simple random search}
\begin{algorithmic}[0]
\Function{simpleRandomSearch}{\textbf{z}, \textbf{X}, d, k}
\State $D_{min} \gets$ 1
\For {$i\gets$ 1 \textbf{to} k}
\State $\boldsymbol{u}\gets$ randomSphere(d) 
\State $D\gets D\left(\boldsymbol{z}\boldsymbol{u}^\top|\boldsymbol{X}\boldsymbol{u}^\top\right)$
\State $D_{min} \gets \min(D,D_{min})$
\EndFor
\State \Return $D_{min}$
\EndFunction
\end{algorithmic}
\label{alg:simpleRandomSearch}
\end{algorithm}

\medskip
\paragraph{Refined random search.}
\medskip

Refined random search (RRS) is an enhanced variation of RS that leverages refinements to concentrate the search direction within specific regions. This method involves partitioning the total number of search directions $k$ into $r$ distinct refinements, where each refinement is performed using $m=\lceil\frac{k}{r}\rceil$ directions. The underlying concept of RRS is to perform depth computation through a series of iterative steps. This process employs a set of procedures to identify and select a region of interest, subsequently generating search directions within the relevant subspaces. By focusing the search effort in targeted regions, RRS aims to improve the efficiency and effectiveness of the optimization process.


\medskip

The first step is performed using RS, with a focus on a hemispherical cap.  It is important to note that for this purpose, the depth function satisfies the property $D(\boldsymbol{z}\boldsymbol{u}^\top|\boldsymbol{X}\boldsymbol{u}^\top)=D(-\boldsymbol{z}\boldsymbol{u}^\top|-\boldsymbol{X}\boldsymbol{u}^\top)$.  

\medskip

At each refinement, the depth is computed with respect to a set of directions $\boldsymbol{U}=(\boldsymbol{u_1},\dots,\boldsymbol{u_m})$, which are generated around a central point referred to as the pole. Among the $m$ mapped depth values, the lowest value is selected as the current minimal depth. The corresponding direction $\boldsymbol{u}$ is associated with this minimal depth is then designated as the new pole for subsequent iterations.

\medskip

This selected direction serves as the center of the region in which the subsequent mapping is conducted. The size of the explored region is decreased by a factor $\alpha$, a process referred to as spherical cap shrinkage. A new set of directions is then generated within this smaller spherical cap, and the entire process is repeated $r$ times. During each refinement, a new pole is identified based on the direction corresponding to the lowest depth value. 



\medskip

This approach can be decomposed into granular logical functions, each serving specific purposes, in alignment with our simple function approach. The primary steps for RRS are described in Algorithm \ref{alg:RefRandSearch}. It is important to note that certain arguments have been modified in comparison to the random search (RS) presented in Algorithm \ref{alg:simpleRandomSearch}. In this context, the number of directions $k$ denotes the total number of directions considered across all refinements combined, where $m$ is the number of directions in each of the $r$ refinements. Spherical shrinkage refers to the factor by which the size of the spherical cap is reduced ($\epsilon$), initialized as $\frac{\pi}{2}$.

\begin{algorithm}[!h]
\footnotesize
\caption{Computing depth with Refined Random Search}
\begin{algorithmic}[0]
\Function{refinedRandomSearch}{$\boldsymbol{z},\boldsymbol{X},k, r, \alpha,d$}
\State $\boldsymbol{p}\gets \boldsymbol{e_1}$
\State $m \gets \lceil \frac{k}{r}\rceil$
\State $\epsilon \gets \frac{\pi}{2}$ 
\State $D_{min} \gets 1$
\For {$l \gets 1$ \textbf{to} $r$}
\For {$i\gets$ 1 \textbf{to} $m$}
\State $\boldsymbol{u}\gets$ randomSpherePole($\boldsymbol{p}, \epsilon$) 
\State $D\gets D(\boldsymbol{z}\boldsymbol{u}^\top|\boldsymbol{X}\boldsymbol{u}^\top)$ 
\If {$D < D_{min}$}
\State $D_{min}$ $\gets$ $D$
\State $\boldsymbol{u_j}\gets \boldsymbol{u}$
\EndIf
\EndFor
\State $\boldsymbol{p}\gets \boldsymbol{u_j}$
\State $\epsilon \gets \epsilon\cdot\alpha$
\EndFor
\State \Return $D_{min}$
\EndFunction
\end{algorithmic}
\label{alg:RefRandSearch}
\end{algorithm}

Generating random directions within the search region is not as straightforward as in RS direction generation. To account for region constraints, Algorithm \ref{alg:randomSphere} is modified with a series of reflections and transformations, as outlined in Algorithm \ref{alg:randomSpherePole}, to map the direction within the desired region.
\medskip

\begin{algorithm}[!h]
\footnotesize
\caption{Creating a random direction on the unit sphere within the search region}
\begin{algorithmic}[0]
\Function{randomSpherePole}{$\boldsymbol{p},\epsilon,d$}
\State $\boldsymbol{u}$ $\gets0_d$
\State $\boldsymbol{u}_1$ $\gets cos(\mathcal{U}([0,1])\cdot\epsilon)$
\State $\boldsymbol{u}_{2:d}$ $\gets \sqrt{1-u_1^2}$ $\cdot$ randomSphere(d-1)
\State $\lambda \gets (\boldsymbol{p}\boldsymbol{u}^\top-\boldsymbol{u_1})\mathbin{/} (1-\boldsymbol{p_1})$ 
\State $\boldsymbol{u_1}=\boldsymbol{u_1}+\lambda$
\For{i $\gets$ 0 \textbf{to} d}
\State$\boldsymbol{u_i}=\boldsymbol{u_i}-\lambda$
\EndFor
\State \Return $\boldsymbol{u}$
\EndFunction
\end{algorithmic}
\label{alg:randomSpherePole}
\end{algorithm}

\section{Depth convergence and illustration}
\label{sec:Depth}

\begin{figure}[ht]
    \centering
    \includegraphics[width=0.99\linewidth]{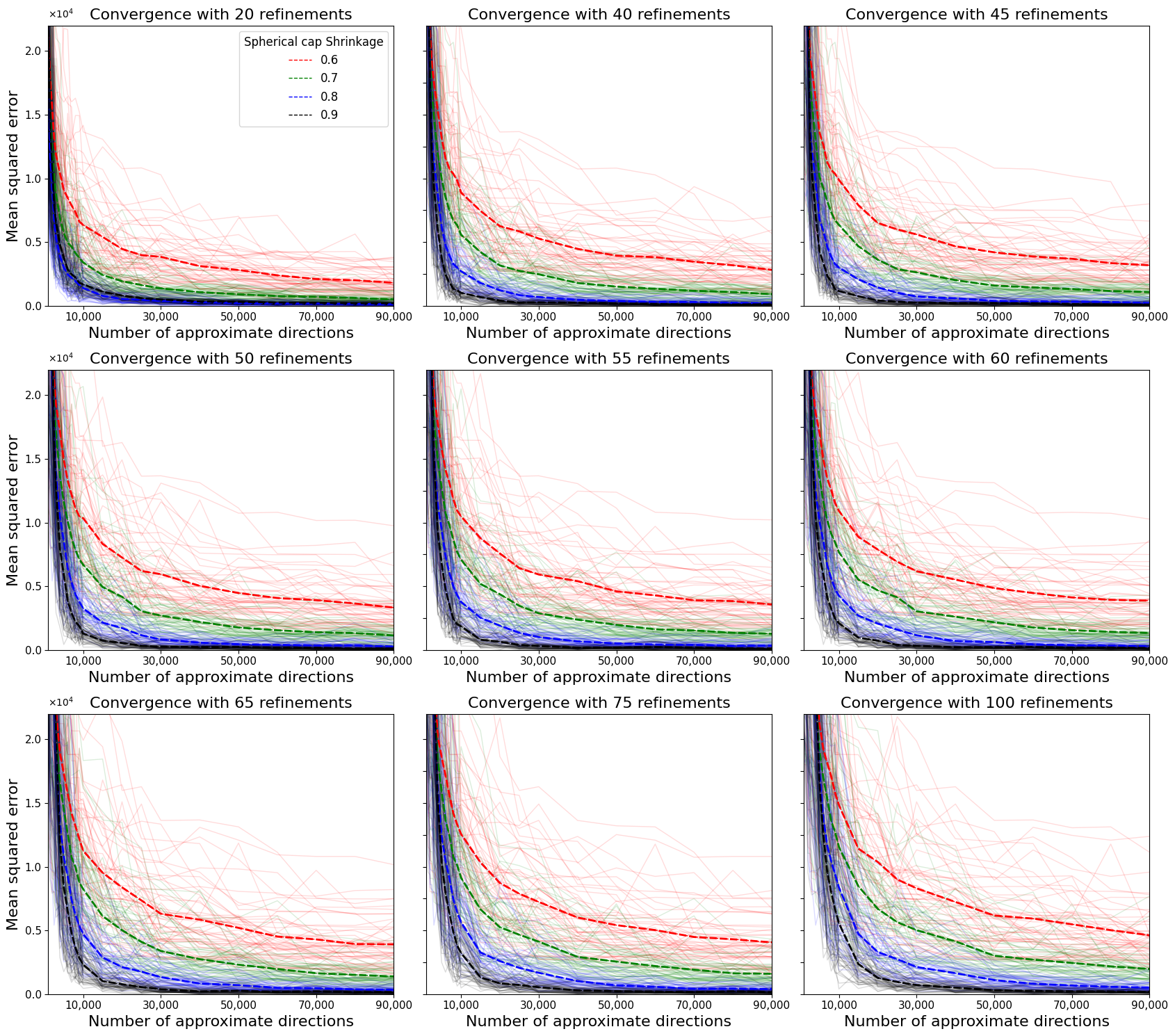}
    \caption{Convergence study using MSE for $D_P$ with respect to a grid of refinements, directions, and spherical shrinkage values, with $d=50$. The number of refinements takes value $r\in\{20,40,45,50,55,60,65,75,100\}$; the number of directions range from 200 to 90,000 and $\alpha \in\{0.6,0.7,0.8,0.9\}$. The data set size is fixed at 50,000. MSE is computed for 50 random points from $\boldsymbol{X}$ with respect to a reference value.}
    \label{supp-fig:conv20_40a100}
\end{figure}

The convergence of depth to its minimum value is analyzed using a multivariate normal distribution, as detailed in Section \ref{sec:fineTun} of the paper. Figure \ref{supp-fig:conv20_40a100} illustrates this convergence for 50 sample points, examining the relationship between spherical cap shrinkage ($\alpha$) - represented by different colored lines - and the number of directions. The space dimension is fixed at d=50. Each subplot corresponds to a different number of refinements, ranging from 20 to 100. The results consistently demonstrate that higher values of ($\alpha$) lead to faster convergence, irrespective of the number of refinements employed.

\begin{figure}[ht]
    \centering
    \includegraphics[width=1\linewidth]{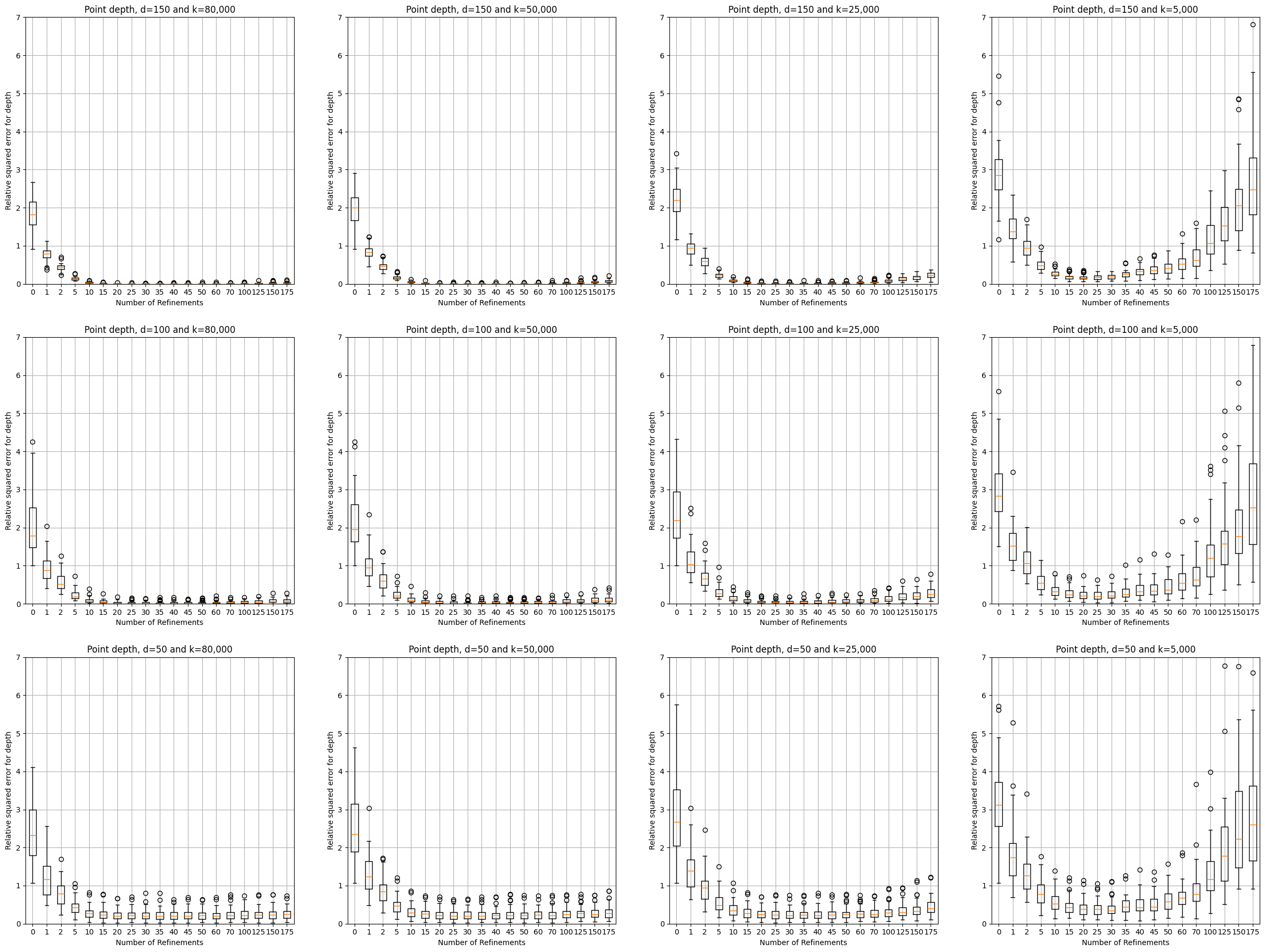}
    \caption{Relative squared error depth box-plot for projection depth. Computation takes into consideration a data set with 10,000 points generated from a multivariate normal distribution centered in $0_d$ and a Toeplitz matrix for the covariance, as described in Section \ref{sec:fineTun} of the main text. From this data set, 50 points are selected for depth analysis.}
    \label{supp-fig:bp_point}
\end{figure}

The relative squared error of computed projection depth values is presented in Figure \ref{supp-fig:bp_point}, comparing them against minimum computed values. The analysis examines various combinations of space dimensions, refinement quantities, and direction counts. The findings reveal that the computation exhibits greater sensitivity to refinement levels when fewer approximate directions are used.

\section{Graphical processing unit parallel programming}
\label{sec:GPU}

The objective of parallel computation is to expedite independent calculations by harnessing the power of multiple CPU processors or, as utilized in this article, multi-core graphical processing units (GPUs). Transitioning from a CPU-based to a GPU-based program is not a trivial task; several critical steps are required for an application to operate effectively in such a multi-core and multi-threaded environment. Key considerations include thread and device synchronization, memory allocation, cached memory access, and flow divergence. Each of these aspects must be meticulously addressed to ensure optimal performance and efficiency in a parallel computing framework.

\medskip

Calculations performed on the graphics card utilize a multithreaded approach, wherein computations and functions are decomposed into smaller instructions known as threads. Each thread is constructed based on a specific task assigned to the computer, referred to as the (\textit{kernel code}). Threads are organized and grouped together, typically in sets of 32, forming a unit known as a \textbf{warp}. A collection of warps constitutes a \textbf{threadblock}. A collection of threadblocks is called a grid, and threadblocks within the same grid are allocated and executed in a \textbf{Streaming Multiprocessor} (SM). 

\medskip

This architecture is constructed and specified using an execution configuration that delineates the number of blocks per grid and threads per block. At a lower level, the kernel code incorporates predefined variables, namely \textit{blockId} and \textit{threadId}. These variables are instrumental in efficiently monitoring the computations being executed in various sections of the GPU, thereby mitigating potential asynchronicity issues \citep{ryoo2008optimization,nickolls2010gpu}.

\medskip

In addition to the extra considerations that require computational time, it is important to note that GPU cores operate at a lower frequency, measured by the number of cycles executed in a given time period, compared to CPU cores. Consequently, individual computations tend to be slower on graphics units due to factors such as memory allocation and lower core performance. However, graphics units compensate for this by leveraging a significantly higher number of cores, typically numbering in the thousands. In contrast, CPUs usually feature dozens of cores. This architectural difference allows GPUs to excel in parallel processing tasks, despite the slower performance of individual cores.

\medskip

Moreover, the necessity to break down operations into simpler tasks mandates the creation of threads with straightforward commands. These parallel computations are particularly well-suited for floating-point operations, such as matrix multiplication. Given these advantages and limitations, the creation and execution of threads must be meticulously coordinated, especially for large-scale applications. In extensive computational routines, leveraging parallel multithreading can result in substantial time savings, effectively addressing one of the most significant challenges in depth computation.

\medskip

Depth computation can significantly benefit from parallel approaches not only in data set projection but also in other operations. Generating random directions and computing univariate depth, although on a smaller scale, can also leverage these methods effectively. A simplified overview of the complete Refined Random Search (RRS) process is presented in the pseudocode \ref{alg:GeneralParallelRRS}. More intricate steps are involved in random generation and univariate depth operations. These include median computation for univariate $D_P$ calculation and conditional comparisons for $D_H$. These more extensive computations are particularly susceptible to flow divergence, necessitating special attention to ensure optimal performance.

\begin{algorithm}[!h]
\footnotesize
\caption{Simple view of parallel computation of refined random search}
\begin{algorithmic}[0]
\Function{GeneralParallelRRS}{$\boldsymbol{X},\boldsymbol{z},r,k,\alpha,d$}
\State \textit{\textbf{Initialisation} }
\State $D_m\gets1$
\State $\epsilon\gets\frac{\pi}{2}$
\State $m\gets\lceil\frac{k}{m}\rceil$
\State $\boldsymbol{U}$ $\gets0_{m\times d}$
\State $\boldsymbol{p}$ $\gets \boldsymbol{e_1}$
\For{$l\gets 1$ to $r$}
\State \textit{\textbf{Step 1:} Generation}
\State $\boldsymbol{U}$ $\gets DirMassive(p,m,\epsilon)$
\State \textit{\textbf{Step 2:} Projection}
\State $\boldsymbol{P}^{(\boldsymbol{X})}$ $\gets ProjMassive$($\boldsymbol{X},\boldsymbol{U},\kappa$)
\State $\boldsymbol{P}^{(\boldsymbol{z})}$ $\gets ProjMassive$($\boldsymbol{z},\boldsymbol{U},\kappa$)
\State \textit{\textbf{Step 3:} Computation}
\State $\boldsymbol{D} \gets UniDepthMassive(\boldsymbol{P}^{(\boldsymbol{z})},\boldsymbol{P}^{(\boldsymbol{z})},notion)$ \Comment{Univariate depth depends on chosen notion}
\State $D_c\gets \min(\boldsymbol{D})$ \Comment{$D_c$ corresponds to the current minimal depth}
\State $\epsilon\gets\epsilon\cdot\alpha$
\If{$D_c<D_m$}
\State $D_m\gets D_c$
\State $\boldsymbol{p}\gets U_c$ \Comment{$U_c$ is the direction corresponding to $D_c$}
\EndIf
\EndFor
\State \Return $\boldsymbol{D_m}$
\EndFunction
\end{algorithmic}
\label{alg:GeneralParallelRRS}
\end{algorithm}

\end{document}